\documentclass[reprint,showpacs,preprintnumbers,amsmath,superscriptaddress,amssymb,pre]{revtex4-1}
\usepackage{graphicx,wasysym}
\usepackage{dcolumn}
\usepackage{bm}
\usepackage{bbm}
\usepackage{natbib}
\usepackage{tabulary}
\usepackage{color}

\usepackage{color} 
\usepackage[normalem]{ulem}

\usepackage{graphicx}
\usepackage{amsmath}
\usepackage{color}
\usepackage{float}
\usepackage{bm}

\newcommand{\Dt}{D_\mathrm{t}}
\newcommand{\Dr}{D_\mathrm{r}}

\newcommand{\bbf}[1]{\boldsymbol{#1}}

 \usepackage[T1]{fontenc}

\begin{document}

\title{Entropy production fluctuations encode collective behavior in active matter}

\author{Trevor GrandPre}
\affiliation{Department of Physics, University of California, Berkeley CA 94609}
\author{Katherine Klymko}
\affiliation{Computational Research Division, Lawrence Berkeley National Laboratory CA 94609}
\author{Kranthi K. Mandadapu}
\affiliation{Department of Chemical and Biomolecular Engineering, University of California, Berkeley CA 94609}
\affiliation{Chemical Science Division, Lawrence Berkeley National Laboratory, Berkeley, CA 94609}
\author{David T. Limmer} 
\affiliation{Chemical Science Division, Lawrence Berkeley National Laboratory, Berkeley, CA 94609}
\affiliation{Department of Chemistry, University of California, Berkeley CA 94609}
\affiliation{Materials Science Division, Lawrence Berkeley National Laboratory, Berkeley, CA 94609}
\affiliation{Kavli Energy NanoScience Institute, Berkeley, CA 94609}
\email{dlimmer@berkeley.edu}

\date{This manuscript was compiled on \today}

\date{\today}
\begin{abstract}
We derive a general lower bound on distributions of entropy production in interacting active matter systems. The bound is tight in the limit that interparticle correlations are small and short-ranged, which we explore in four canonical active matter models. In all models studied, the bound is weak where collective fluctuations result in long-ranged correlations, which subsequently links the locations of phase transitions to enhanced entropy production fluctuations. We develop a theory for the onset of enhanced fluctuations and relate it to specific phase transitions in active Brownian particles. We also derive optimal control forces that realize the dynamics necessary to tune dissipation and manipulate the system between phases.  In so doing, we uncover a general relationship between entropy production and pattern formation in active matter, as well as ways of controlling it.
\end{abstract}


\maketitle

Active matter systems are defined by forces that inject energy locally into individual particles, driving nonequilibrium steady-states that continuously dissipate energy.
This persistent dissipation and its associated entropy production have been shown to have deep connections with structural and dynamic properties of active matter \cite{nardini2017entropy,fodor2020dissipation,del2018energy,gaspard2018fluctuating,dasbiswas2018topological,soni2019odd,liao2020rectification,PhysRevLett.124.240604,fodor2016far,mandal2017entropy}. Subsequently, understanding the contributions to the entropy production in active matter is the first step in manipulating their emergent order \cite{metselaar2019topology,li2019shape,wang2019shape,morris2019active,bain2019dynamic,morin2018flowing,bricard2013emergence,mietke2019minimal,morozov2017chaos}, designing active metamaterials with novel responses \cite{souslov2017topological,ropp2018dissipative,takatori2020active,chaudhuri2014active}, and utilizing active heat engines \cite{PhysRevX.9.041032,krishnamurthy2016micrometre,zakine2017stochastic,martin2018extracting,saha2018stochastic,PhysRevE.102.010101,chaki2018entropy}.
Stochastic thermodynamics provides a framework for studying entropy production and has supplied general theories that constrain its statistics \cite{jarzynski1997nonequilibrium,crooks1999entropy,gallavotti1995dynamical,gaspard2017communication,horowitz2019thermodynamic} and its role in nonequilbrium response   \cite{di2018kinetic,owen2020universal,nardini2018process,gao2019nonlinear,dechant2020fluctuation,polettini2019effective,barbier2018microreversibility,epstein2020time,hargus2020time,wagner2019response,asheichyk2019response,dal2019linear,caprini2018linear,merlitz2018linear,liao2019mechanism}.
Here, we provide a general bound on the distributions of entropy production for interacting active matter using stochastic thermodynamics and large deviation theory~\cite{chetrite2015nonequilibrium}. 
While not universal like the thermodynamic uncertainty principle~\cite{barato2015thermodynamic,PhysRevLett.116.120601}, the specific consideration of active matter admits a tight bound generically, and one in which deviations can be physically understood. 
The bound we present is valid arbitrarily far from equilibrium for self-propelled particles and  is saturated in the limit that the interparticle contribution to the entropy production is small. Near phase transitions, the bound is weak as fluctuations are enhanced due to emergent effective long-ranged interactions that we quantify. This work provides a link between entropy production fluctuations and collective phenomena in active matter. 

\begin{figure*}[t]
\includegraphics[width=\textwidth]{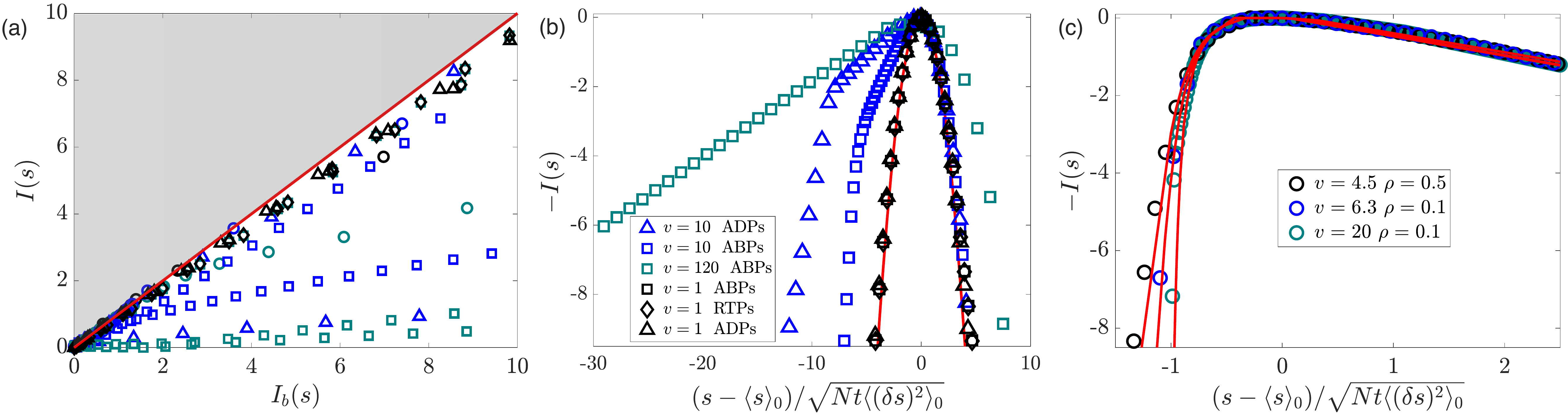}
\caption{Entropy production fluctuations for a variety of active matter systems. (a) Rate function obtained by importance sampling versus the bound in Eq.~\ref{boundI} with the symbols corresponding to the labels in (b) and (c).  (b) Entropy production fluctuations for $\rho =0.1$,  ABPs with $\Dr=3$, ADPs with spring constant $k=100\ \epsilon/\sigma^2$ and rest length $l=1.5\sigma$, and RTPs with a tumble rate $\gamma=1$ for different self-propulsion values. (c) Entropy production fluctuations for AOUPs for different parameter ranges and $\Dr=1$. 
In all panels the red line denotes $I_b(s)$ and the errorbars are smaller than the symbols.}
\label{Fi:1}
\end{figure*}

We consider active matter systems that are self-propelled and whose equations of motion are of the form 
\begin{equation}
    \bbf{\dot{r}}_i=v\,\bbf{b}_i+\mu\bbf{F}_i \left(\bbf{r}^N \right)+\sqrt{2 \Dt}\bbf{\eta}_i \ ,
    \label{dynamics}
\end{equation}
where $\bbf{r}_i$ denotes the position of the $i$'th particle, $v$ and $\bbf{b}_i$ set the typical magnitude and direction of self-propulsion, $\mu$ is a single particle mobility, and $\bbf{\eta}_i$ is a Gaussian white noise with $\langle \eta^\alpha_i(t)\rangle =0$ and $\langle \eta^\alpha_i(t)\eta^\beta_j(t')\rangle = \delta_{ij} \delta_{\alpha,\beta} \delta(t-t')$ for the $\alpha$ and $\beta$ components of the random force. The translational diffusion coefficient,  $\Dt$, satisfies a fluctuation-dissipation relation, $\Dt=\beta^{-1} \mu $ where $\beta^{-1}$ is the temperature times Boltzmann's constant. Throughout, we take $\mu=\Dt=1$. The interparticle forces are conservative, $\bbf{F}(\bbf{r}^N)=-\nabla U(\bbf{r}^N)$, and in general depend on all $N$ particles' positions, $\bbf{r}^N$. This class of active matter has a non-conservative self-propulsion term, $v \bbf{b}$, which is driven by a constant energy supply. Our formulation is independent of the statistics and dynamics of the self propulsion vector, $\bbf{b}$, and may be correlated due to aligning interactions. The dynamics of the orientation vector $\bbf{b}$ are model specific and discussed in Appendix \ref{app:simulation_details}, however our results are largely independent of its form. For concreteness, below we will consider collections of interacting active Brownian particles (ABPs), active dumbbells (ADPs), run and tumble particles (RTPs), and active Ornstein-Uhlenbeck particles (AOUPs). 

The entropy production follows from  time reversal symmetry arguments of stochastic thermodynamics~\cite{lebowitz1999gallavotti,seifert2005entropy,maes2003time,caprini2019entropy,chaki2019effects},
$\Delta S = \ln P[\Gamma]/P[\tilde{\Gamma}]$, where $P[\Gamma]$ is the probability of a forward trajectory $\Gamma=\{\bbf{r}^N(t),\bbf{b}^N(t)\}$ and $P[\tilde{\Gamma}]$ is the probability of observing the time-reversed trajectory.
We use the convention that the parameter $v$ is even under time-reversal consistent with previous work~\cite{cagnetta2017large,nemoto2019optimizing, shankar2018hidden,szamel2019stochastic,pietzonka2017entropy}. 
This convention ensures that there is a nonvanishing entropy production in the limit of noninteracting particles with no external fields accounting for the energy injected into the single particles to drive persistent motion. The choice of convention for time reversal  without an underlying microscopic model of self-propulsion is somewhat arbitrary~\cite{dabelow2019irreversibility}. 
However, all of the collective phenomena reported below are independent of the convention~\cite{speck2017stochastic}. 
The convention we follow is analogous to the active work~\cite{shankar2018hidden} and can be derived independently from mechanical considerations.
Under this time reversal convention in the long time limit, the entropy production is
\begin{equation}
    \Delta S=\frac{v}{\Dt}\sum_{i=1}^N \int_0^t dt'   \  \bbf{b}_i\circ\bbf{\dot{r}}_i \ \ ,
    \label{EP}
\end{equation}
where $\circ$ denotes a Stratonovich product (see Appendix \ref{app:EP}). This definition codifies the amount of energy directly translated into motion in the form of persistent displacement~\cite{speck2018active,pietzonka2017entropy,dabelow2019irreversibility}. 

One convenient way to characterize the statistics of $\Delta S$ is through its scaled cumulant generating function (CGF). For the time and system size intensive entropy production, $s=\Delta S/(N t)$, the CGF is defined as,
\begin{equation}
\psi(\lambda)=\frac{1}{tN}\ln\left< e^{\lambda  s(\Gamma) N t} \right>_0,
\label{CGF}
\end{equation}
where $\langle \dots \rangle_0$ denotes average over paths and  $\lambda$ is the counting variable that probes rare fluctuations of the entropy production when nonzero. Cumulants of the entropy production are computable from $\psi(\lambda)$ through derivatives with respect to $\lambda$. We define a rate function 
\begin{equation}
I(s)=-\frac{1}{Nt} \ln \left< \delta [s- s(\Gamma) ] \right>_0 \, ,
\end{equation}
where $\delta(s)$ is Dirac's delta function. The rate function is the logarithm of the probability of $s$  scaled by time and particle number. We are interested in the fluctuations of $s$ in the macroscopic limit at long time and large system size, where $I(s)$ can be calculated by a Legendre-Fenchel transform,
$I(s)=\max_{\lambda}\left[\lambda s-\psi(\lambda)\right]$. The transient fluctuations of entropy production would require an alternative method from the Legendre-Fenchel transform and are not considered in our work. 

Calculating $\psi(\lambda)$ or $I(s)$ exactly for interacting systems is difficult because of many-body correlations. However, we find that $\psi(\lambda)$ can generally be rewritten by factoring out the single particle part,
\begin{equation}
\label{eCGF}
\psi(\lambda) = \psi_f(\lambda) + \frac{1}{N t} \ln \left \langle e^{\lambda \Delta W} \right \rangle_{\bbf{u}_{\lambda}} \ ,
\end{equation}
where $\psi_f(\lambda)$ is the CGF for an isolated active particle. The remaining contribution to $\psi(\lambda)$ represents interparticle correlations and is given by the CGF of
\begin{equation}
\Delta W= \beta  v \sum_{i=1}^{N} \int_0^{t} dt' \,  \bbf{b}_i \cdot  \bbf{F}_i \ ,
\end{equation}
averaged over an ensemble with an additional force  $\bbf{u}_{\lambda}$. The force $\bbf{u}_{\lambda}$ is the optimal control force to realize rare entropy production fluctuations for an isolated particle and its model specific form is considered below. The observable $\Delta W$ is the dimensionless work done on the surrounding particles due to self-propulsion. By applying Jensen's inequality to Eq.~\ref{eCGF},  
\begin{equation}
\label{bound}
\psi(\lambda) \ge \psi_f(\lambda) + \beta \lambda v \left \langle \bbf{b} \cdot  \bbf{F} \right \rangle_{\bbf{u}_{\lambda}} \ ,
\end{equation}
$\psi(\lambda)$ is bounded (see Appendix \ref{app:bounds}). The correction over the single particle CGF can be interpreted as $\beta \lambda v$ times the effective drag a tagged particle feels in the direction of the self-propulsion due to the surrounding particles \cite{speck2015dynamical}. This gives rise to an effective velocity that is smaller than $v$ and dependent on the density and $\lambda$ \cite{grandpre2018current}.

Inserting the bound on the CGF in Eq.~\ref{bound} into the Legendre-Fenchel transform, we derive a bound on the distribution of the entropy production, $I_b(s)$, 
\begin{equation}
\label{boundI}
I(s)\le I_b(s) =\max_{\lambda} \left[\lambda s-\psi_f(\lambda) - \beta \lambda v \left \langle \bbf{b} \cdot  \bbf{F} \right \rangle_{\bbf{u}_{\lambda}} \right ].
\end{equation}
By construction the bound recovers the correct mean dissipation and is tight far into the tails of the distribution in the limit that fluctuations in $\Delta W$ are small and the saddle point approximation to its CGF is accurate. Data in Fig.~\ref{Fi:1} confirms the upper bound for all of the active matter models studied. Throughout, $I(s)$ is computed using the cloning algorithm \cite{giardina2006direct,ray2018exact} and $I_b(s)$ by computing $\left \langle \Delta W \right \rangle_{\bbf{u}_{\lambda}}$ from direct simulations. All simulations are done with a WCA interparticle potential~\cite{weeks1971role}, 
$
U(r)=4\epsilon \left [\left ( \frac{\sigma}{r} \right)^{12}-\left ( \frac{\sigma}{r} \right)^{6} \right ]+\epsilon$ for $r\le2^{1/6}\sigma$ and zero otherwise. The parameter $\epsilon$ is the energy scale of the interactions and $\sigma$ is the particle diameter. ADPs have an added harmonic potential between composite particles. Our results are presented with a non-dimensional $v$ in units of $\Dt/\sigma$, $\gamma$ and $\Dr$ in units of $\Dt/\sigma^2$, and bulk density $\rho$ in units of $1/\sigma^2$ in two dimensions. Also, $\Dt$, and $\beta$ are set to 1. Data in Fig.~\ref{Fi:1}a shows that there are large parameter regimes where the bound is tight.  In practice, the bound is accurate when the system is away from dynamical phase transitions, this is valid when $\rho (v/\Dr)^2<1$. Nevertheless, even when $\rho (v/\Dr)^2 \approx 1$ we find the bound is still reasonably tight.

The detailed forms for $I(s)$ and $I_b(s)$ are distinct for different models of active matter. For  ABPs,  ADPs, and RTPs, the entropy production fluctuations are Gaussian for isolated particles, with $\psi_{f}(\lambda)=v^2\lambda(1+\lambda)/\Dt$ (see Appendix \ref{app:free}).
The corresponding control force, $\bbf{u}_{\lambda}=2\lambda v\bbf{b}$,
 is appended to the existing forces in Eq.~\ref{dynamics} such that rare entropy production fluctuations are realized by a renormalized velocity, $v_\lambda= v(1+2 \lambda)$. This $\psi_f(\lambda)$ gives rise to a bound that is nearly Gaussian, as shown in Fig.~\ref{Fi:1}b. For low densities and low velocities, $I(s)\approx I_b(s)$. Increasing $v$, the bound weakens for smaller than average entropy production fluctuations, $s <\langle s\rangle_{0}$. 
Fluctuations that result in larger than average entropy production, $s \gtrsim \langle s\rangle_{0}$, for large $v$ are more probable than predicted by the bound due to neglecting contributions from interparticle correlations. However, the relative error between the entropy production distribution and the bound decreases into the tails due to the increasingly independent particle behavior elaborated upon below.

For isolated AOUPs, the entropy production fluctuations are generically non-Gaussian and  
$
\psi_f(\lambda) = \Dr \left ( 1- \sqrt{1-2 v^2 \lambda(1+\lambda)/\Dt \Dr } \right )
$, where $\Dr$ is the rotational diffusion constant  
(see Appendix \ref{app:free}). The fluctuations in $s$ are Gaussian only near the mean and are asymmetric \cite{shankar2018hidden}.
 This is in contrast to the Gaussian distribution that would be predicted by the thermodynamic uncertainty relations, and reflects the finite memory in the self-propulsion vector.
 The control force includes the same renormalized velocity as for ABPs, but in addition includes a force on the particle's orientation, $\bbf{u}_{\lambda} = \psi_f(\lambda)\bbf{b}$. In Fig. \ref{Fi:1}c, we see that the bound gives an accurate prediction of the fluctuations across the densities and $D_\mathrm{a}$'s considered. The fluctuations are still enhanced relative to the bound for  $s <\langle s\rangle_{0}$, though less so than in Fig.~\ref{Fi:1}b.
\begin{figure}[t]
\includegraphics[width=8.5cm]{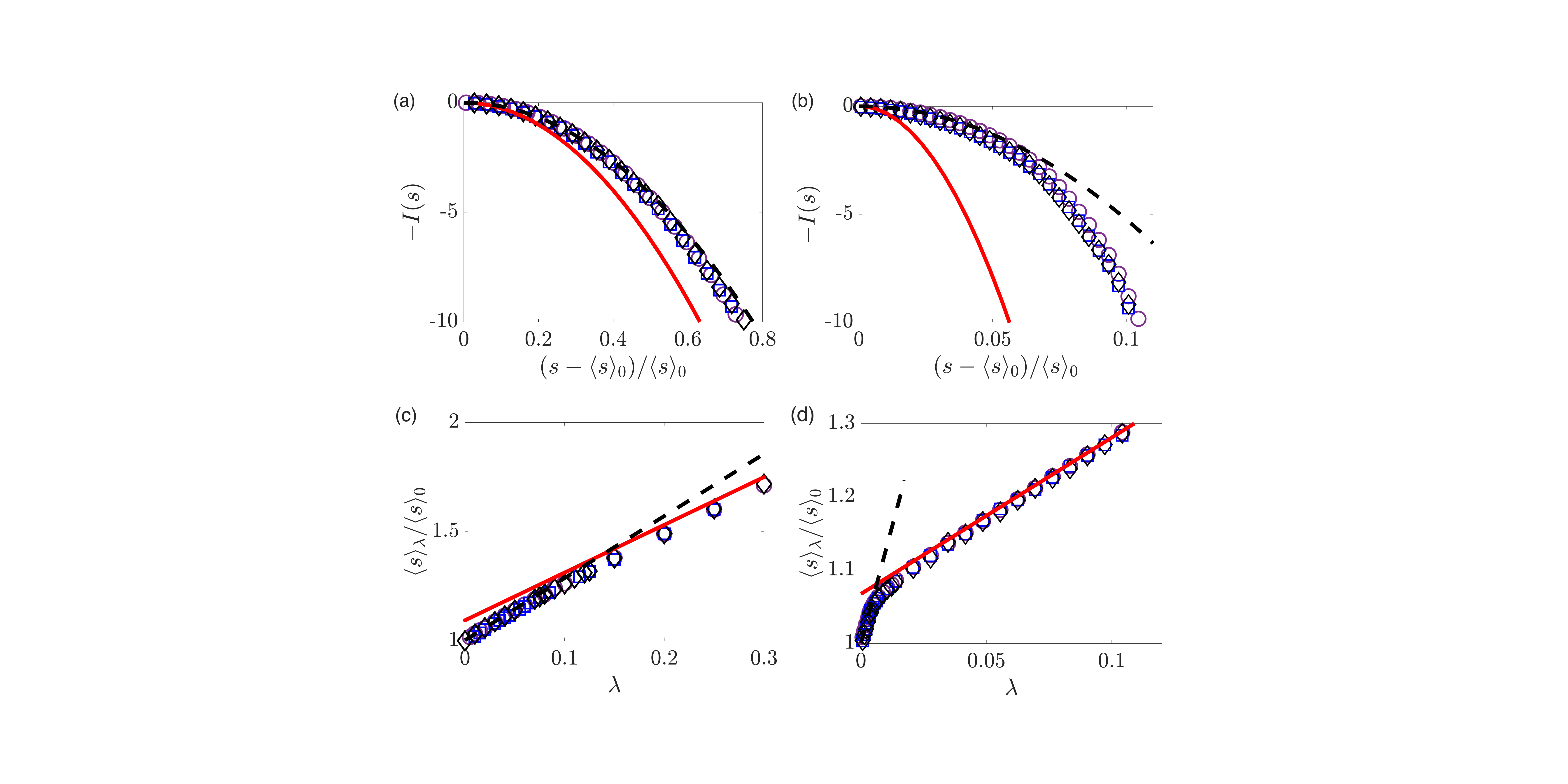}
\caption{Larger than average entropy production fluctuations for ABPs with $N=10$ (purple circles), $20$ (blue squares), and $40$ (black diamonds). 
Distribution of entropy production for a) $v=10$ and b) $v=120$ with $\rho=0.1$. In a) and b), the red lines are $I_b(s)$ and the dashed black lines are fits at $\lambda=0$ to extract the second cumulant. The average entropy production at finite $\lambda$ for c) $v=10$ and d) $v=120$ with $\rho=0.1$. The dashed lines are from the cumulant fits in a) and b), and the red line is the non-interacting rate function. 
}
\label{Fi:2}
\end{figure}

In order to understand the origins of the deviations from the bound and the connections to collective behavior in active matter, we consider in detail a system of ABPs at conditions near and far from its motility induced phase separation (MIPS) transition.  Additionally, the asymmetry of entropy production fluctuations about its average, motivates us to consider separately fluctuations of $s>\langle s \rangle_0$ and $s<\langle s \rangle_0$. In Fig. \ref{Fi:2}, the distributions for $s>\langle s \rangle_0$ are shown for $v=10$ and $v=120$, for a variety of system sizes at fixed density, $\rho=0.1$. While the probability is larger than predicted by the bound, it can be perturbatively corrected. Specifically, we can expand Eq.~\ref{eCGF} up to the second cumulant,  $\psi(\lambda) \approx \psi_f(\lambda) + \left (\lambda \left \langle \Delta W \right \rangle_{0} + \lambda^2 \left \langle \delta \Delta W^2 \right \rangle_{0}/2 \right )/N t$. The result of this approximation to the rate function is shown in Figs. \ref{Fi:2} a) and b). For $v=10$ the fluctuations are well described by the cumulant approximation, while for $v=120$ asymptotic entropy production fluctuations are narrower than predicted. 

The asymptotic behavior for $s \gg \langle s\rangle_{0}$ is well described by free particle motion for all $v$'s. This can be seen by considering $d \psi/d\lambda = \langle s \rangle_\lambda$  from 
$\langle s \rangle_\lambda = \int ds\, s \exp \{N t [-I(s)+\lambda s- \psi(\lambda)]\}$, which is a direct probe of the tails of $I(s)$. As shown in Figs. \ref{Fi:2} c) and d), for both large and small $v$, $\langle s \rangle_\lambda$ exhibits a crossover from Gaussian statistics. Near $\lambda=0$, $\langle s \rangle_\lambda$ varies linearly with $\lambda$ with a slope given by the variance $\langle \delta s^2\rangle_0$. For $\lambda \gg 0$, $\langle s \rangle_\lambda$ varies linearly with $\lambda$ with a slope given by the free particle variance. An analogous crossover has been noted in the current statistics of an interacting tagged ABP \cite{grandpre2018current}. The asymptotic free behavior implies that the most likely way for the system to produce large amounts of entropy is to suppress density correlations and decrease $\Delta W$. This behavior results from the system adopting a net orientation for the particles' self-propulsion vector \cite{nemoto2019optimizing, cagnetta2020efficiency}. If the net orientation persists in the thermodynamic limit, it would represent a spontaneous symmetry breaking. 

\begin{figure}[t]
\includegraphics[width=8.5cm]{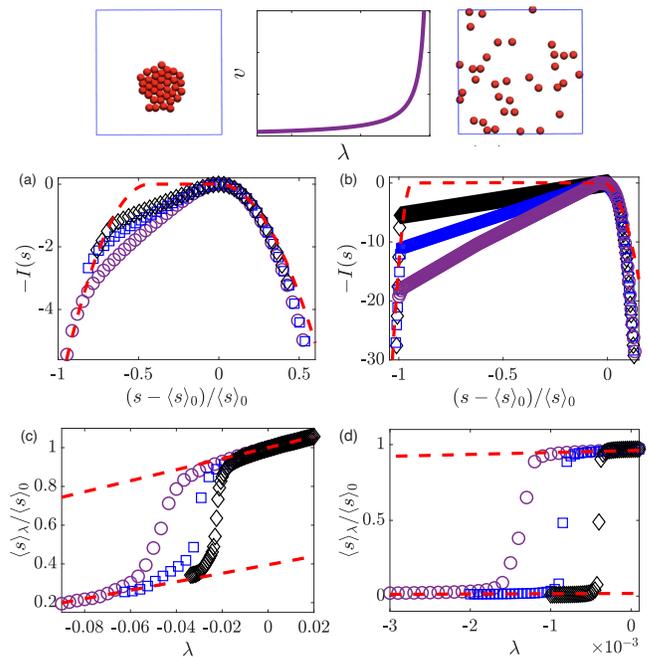}
\caption{
Smaller than average entropy production fluctuations and dynamical phase transition for ABPs for different system sizes $N=10$ (purple circles), $20$ (blue squares), and $40$ (black diamonds). The phase diagram and example structures are illustrated at the top with phase separation on the left of the phase diagram and a homogeneous state on the right. Distribution of entropy production for a) $v=10$ and b) $v=120$ with $\rho=0.1$. In a) and b), the dashed red lines are a Maxwell construction for the dynamical phases. The average entropy production at finite $\lambda$ for c) $v=10$ and d) $v=120$ with $\rho=0.1$. The dashed red lines are from the Gaussian fits in a) and b) used in the Maxwell construction. 
}
\label{Fi:3}
\end{figure}

Fluctuations for  $s<\langle s \rangle_0$ are much larger than predicted by the bound and are collective in origin. 
Fig. \ref{Fi:3} shows the distributions of entropy production and $\langle s \rangle_\lambda$ for $v=10$ and $v=120$ at $\rho=0.1$ for 3 system sizes. The distributions in Figs. \ref{Fi:3} a) and b) show significant finite size effects for $s<\langle s \rangle_0$. In Figs. \ref{Fi:3} c), and d), this is evident by a transition between two types of behavior that sharpens with increasing $N$ and occurs at larger $\lambda$ with increasing $v$ over the limited range of system sizes we can study numerically. These features are a hallmark of a dynamical phase transition, in this case between a dilute phase and a phase separated state reminiscent of MIPS \cite{redner2013structure,cates2015motility,chiarantoni2020work}. As has been found previously~\cite{nemoto2019optimizing}, this shows that the most likely way for the system to produce little entropy is to condense, decreasing the particles' displacement by increasing the effective drag. We find we can describe $I(s)$ by explicitly assuming that each dynamical phase is well approximated by a Gaussian distribution. Specifically, assuming $\psi_i(\lambda)Nt = \lambda \langle \Delta S \rangle_i + \lambda^2 \langle \delta \Delta S^2 \rangle_i /2$ for $i={d,c}$ being the dilute and condensed phases, the rate function can be computed from a contraction principle~\cite{chetrite2015nonequilibrium} for the CGF, $\psi(\lambda)=\max_\lambda [\psi_c(\lambda),\psi_d(\lambda)]$ (see Appendix \ref{app:simulation_details}).  The result is  a Maxwell construction and is shown in Fig.~\ref{Fi:3} to be a good approximation in the infinite system size limit. Due to the exponential difficulty of sampling large deviations in interacting systems we are unable to study larger systems \cite{ray2018importance}. Effects from the relatively large persistent length for $v=120$ may complicate the extrapolation of these finite size effects to larger systems.

For  $s<\langle s \rangle_0$, it is not sufficient to perturbatively correct the bound even for $v=10$, which is far from the MIPS transition. To understand this behavior we have developed a coarse-grained theory. We define a fluctuating density field as $\rho(\bbf{r},t) = \sum_{i=1}^N \delta [\bbf{r}-\bbf{r}_i(t)]$.
With this field, $\Delta W$ can be computed by assuming that the collisions are concentrated directly in front of a tagged particle. Under that assumption, $\Delta W$ can be written in terms of $\rho(\bbf{r},t)$,
$\Delta W \approx -\beta v \int dt  \int d{\bbf{r}} d{\bbf{r}'} \rho(\bbf{r},t) F(|\bbf{r}-\bbf{r}'|) \rho(\bbf{r}',t) /2\, ,
$
which is a convolution of two points of the density field with the interparticle force. For simplicity we have assumed that $F(0)=0$. Further assuming that the force can be Fourier transformed, we find
\begin{equation}
\Delta W=- \frac{\beta v }{2}\int dt  \int d{k} \, | \hat{\rho}(k,t)|^2 \hat{F}(k) \, ,
\end{equation}
where $\hat{\rho}(k,t)$ is the Fourier transformed isotropic density field and $\hat{F}(k)$ the Fourier transformed force.

In order to evaluate the statistics of $\Delta W$, we require an evolution equation for $\hat{\rho}(k,t)$. From the equation of motion for the position and orientation of each particle in the presence of the single particle control force, standard techniques afford an exact equation of motion for $\rho(\bbf{r},t)$ \cite{dean1996langevin}. Its solution is complicated by its non-locality and coupling to a polarization field arising from the dynamics of the particle's orientation (see Appendix \ref{app:density}). Rather than deal with it directly, assuming the system is macroscopically homogeneous on the largest scales, we expect the density field to evolve diffusively. Thus, in the limit that $k \rightarrow 0$, the stochastic equation of motion for $\hat{\rho}(k,t)$ takes the form,
\begin{equation}
\label{OUrho}
\frac{\partial \hat{\rho}(k,t)}{\partial t} \approx -k^2 \mathcal{D}_\lambda  \hat{\rho}(k,t) + \sqrt{2 \Delta_\lambda k^2 } \hat{\eta}_\rho \, ,
\end{equation}
where $\mathcal{D}_\lambda$ is the effective diffusion constant,  $\Delta_\lambda$ is the effective mobility, and $\hat{\eta}_\rho$ is a complex noise \cite{speck2015dynamical,tailleur2008statistical,chakraborti2016additivity,cates2015motility,dolezal2019large}. Assuming that the polarization field relaxes quickly, and linearizing around a homogeneous density, these parameters can be derived explicitly for each active matter model. 

Equation \ref{OUrho} has the form of an independent Ornstein-Uhlenbeck process for each Fourier mode of the density~\cite{dolezal2019large}. The CGF for $\Delta W$ can be solved exactly within this approximate linearized dynamics. Defining $\Delta \psi = \psi-\psi_f$, 
\begin{equation}
\label{deltapsi}
\Delta \psi(\lambda) \approx \frac{1}{N} \sum_{k>0} k^2 \mathcal{D}_\lambda  \left [1-\sqrt{1+\frac{\beta \lambda v_\lambda \hat{F}(k)\Delta_\lambda }{\mathcal{D}_\lambda^2 k^2}} \right ] \, ,
\end{equation}
we get an approximate correction to the bound in Eq. \eqref{bound} due to interparticle correlations. This correction is valid for all positive $\lambda$, but becomes unstable at a critical value $\lambda_c\le 0$ reflecting the breakdown in the linearized evolution equation for $\hat{\rho}(k,t)$. For a finite system with largest wavevector $k=2 \pi \sqrt{{\rho}/N}$, the location of the instability is found by setting the discriminant to zero,
$
\lambda_c \approx -4 \pi^2 \mathcal{D}^2_0 {\rho}/\beta v \hat{F}(0) \Delta_0 N
$
where for the short ranged forces considered, we can approximate the force as $\hat{F}(0)$ and we can neglect the $\lambda$ dependence in $v_\lambda, \mathcal{D}_\lambda$, and $\Delta_\lambda$. 
This instability signals the dynamical phase transition that occurs at $\lambda_c = 0^{-}$ in the thermodynamic limit and whose influence on the dynamics of active matter increases with $v$, and with increasing proximity to MIPS, consistent with the results in Fig.~\ref{Fi:3}. In a phase separated state, $\Delta W$ is a large negative number which counteracts the free particle contribution and reduces the entropy production.

The origin of phase separation can be understood by noting that the optimal control potential which gives rise to rare entropy fluctuations is, for large interparticle separations $r/\sigma\gg 0$ and in the limit $\lambda$ approaches zero (see Appendix \ref{app:density}),
\begin{equation}
   \hat{V}_\lambda \approx -\frac{\beta v\lambda}{N\mathcal{D}_0} \sum_{k>0} \frac{| \hat{\rho}(k,t)|^2  }{2k^2}  \hat{F}(k) \ .
\end{equation}
 The inverse Fourier transform will involve a convolution between the WCA force and $1/k^2$ which gives rise to a Bessel Function. Since the WCA potential quickly decays, the long range contribution in real space is a logrithmic potential,
\begin{equation}
    V_\lambda(r)\approx - \frac{\beta \lambda v}{\mathcal{D}_0}\ln r/2 \ ,
\end{equation}
which is attractive for $\lambda<0$ with a magnitude that depends on $v$ and the control force is $\bbf{u}_\lambda \approx (\beta \lambda v / \mathcal{D}_0) \nabla \ln r/2$. For negative enough $\lambda$ or large enough $v$, this force will give rise to phase separation. This optimal control force is similar to other passive models near diffusive instabilities \cite{das2019variational,tociu2019dissipation,dolezal2019large}. 
 
 The long-ranged effective force demonstrates how effective attractions are introduced by self-propulsion in order to minimize the entropy production. This force is unique and encodes the way in which self-propelled particles interact provided the condition of obtaining a lower than average value of the entropy production. As such, it provides a sharp relationship between entropy production and emergent collective behavior in active matter. Correlations between entropy production and motility induced phase separation have been observed previously at the level of the mean behavior~\cite{crosato2019irreversibility,nardini2017entropy, nigmatullin2019thermodynamic}, however this work codifies that relationship on the level of fluctuations.

For both MIPS and the dynamical transition we discuss, phase separation is the result of a diffusive instability where density accumulates due to unbalanced fluxes made possible by the system being kept from thermal equilibrium. We have shown such collective behavior results from the reduction of entropy production and enhancement of density correlations. Large entropy production by contrast, arises through the suppression of density correlations. Thus, our results show how the structure of entropy production fluctuations are intimately connected to long-ranged correlations in active matter. We expect that deviations from the bound derived here can serve as a guide to identify criticality and novel phases of active matter generally. 

\emph{Acknowledgements}  We thank Chloe Ya Gao for sharing her cloning code that was adapted for our work, John Bell for discussions and computing resources, and Ahmad Omar for helpful discussions. This research used resources of the National Energy Research Scientific Computing Center (NERSC), a U.S. Department of Energy Office of Science User Facility operated under Contract No. DE-AC02-05CH11231. TGP and DTL were supported by  the U.S. Department of Energy, Office of Basic Energy Sciences through Award No. DE-SC0019375. K.K.M was supported by Director, Office of Science, Office of Basic Energy Sciences, of the U.S. Department of Energy under contract No. DEAC02-05CH11231.

\appendix
\section{Simulation Details}
\label{app:simulation_details}

\subsection{Model definitions and parameters}

\subsubsection{ABPs and RTPs}
For both ABPs and RTPs the orientation vector has a fixed magnitude, so in two dimensions it can be uniquely parameterized by an angle $\theta$. For the $i$th particle, $\boldsymbol{b_i}=\{ \cos(\theta_i)\hat{\bbf{x}}_i, \sin(\theta_i)\hat{\bbf{y}}_i \}$, where $\hat{\bbf{x}}_i$ and $\hat{\bbf{y}}_i$ are the unit vectors in the $x$ and $y$ directions, respectively. For ABPs, the dynamics of $\theta_i$ are Brownian,
\begin{equation}
\dot{\theta_i}(t) =  \eta^\theta_i(t)
\end{equation}
where $\eta^\theta_i$ is a Gaussian white noise, satisfying $\langle \eta^\theta_i(t)\rangle =0$ and $\langle \eta^\theta_i(t)\eta^\theta_j(t')\rangle =2 \Dr \delta_{ij} \delta(t-t')$ with $\Dr$ the rotational diffusion constant. We take $\Dr=3 \Dt/\sigma^2$ throughout.

The dynamics of $\theta$ for RTPs are piecewise constant over waiting times, $\tau$, satisfying a Poisson process\cite{solon2015pressure,fily2017equilibrium}. The waiting time distribution is given by an exponential distribution,
\begin{equation}
P(\tau)=\gamma e^{-\gamma \tau},
\end{equation}
with constant reorientation rate $\gamma$. We take $\gamma=\Dt/\sigma^2$. At each $\tau$, the particles reorient by drawing a new $\theta$ chosen uniformly over the range $\left[0,2\pi\right]$.

\subsubsection{ADPs}
Each ADP is composed of two particles that are tethered together by a harmonic bond. The harmonic bond potential is given by $U_\mathrm{H}(r)=k(r-l)^2/2$, where $k$ is the spring constant, $l$ is the rest length, and $r$ is the displacement between the two bonded particles. We take $k=100  \ \epsilon/\sigma^2$ and $l= 1.5\ \sigma$. The self-propulsion direction is along the bond vector. For the $i$th ADP, composed of monomers 1 and 2, 
 $\boldsymbol{b}_i=\hat{\bbf{r}}_{i,12}$ where $\hat{\bbf{r}}_{i,12}$ is the unit displacement vector between monomers 1 and 2. The time evolution of the orientation vector is given by the time evolution of the displacement vector between the two composite particles as dictated by their individual equations of motion \cite{lowen2018active,winkler2016dynamics}.

\subsubsection{AOUPs}
For AOUPs, the self propulsion vector changes both its magnitude and direction. Its equation of motion takes the form of an Ornstein-Uhlenbeck process and given by
\begin{equation}
\boldsymbol{\dot{b}_i}=-D_\mathrm{r}\boldsymbol{b}_i+\bbf{\xi_i}\end{equation}
where $\bbf{\xi_i}$ is a Gaussian random variable satisfying $\langle\xi_{i,\alpha}(t)\rangle=0$ and $\langle\xi^\alpha_{i}(t)\xi^\beta_{j}(t^{\prime})\rangle=\Dr \delta_{ij} \delta_{\alpha, \beta} \delta(t-t^{\prime})$
for each $\alpha,\beta$ component~\cite{fodor2016far,caprini2018linear}.

\subsection{Bound and Cloning calculation details}

 For all simulations we used $N=$10, 20, or 40 particles. A 2-dimensional square box of length $L$ with periodic bounds was used and the length chosen to give the desired density through the equation $L=\sqrt{N/\rho}$.
The equations of motion are discretized using a first order Euler method. Calculations of the rate functions, $I(s)$, require enhanced sampling techniques in order to probe rare fluctuations. For this we use the cloning algorithm\cite{giardina2006direct}. Cloning results were run with 2.4$\times 10^4$-1.5$\times 10^7$ walkers. The cloning parameters varied for the models considered. For ABPs, RTPs, and ADPs, we used a time step of $\delta t=10^{-3}-10^{-5}$, depending on the $v$, a branching time of $t_\mathrm{int}=50\delta t$, and an observation time of $t=10t_\mathrm{int}$. For AOUPs, we used a timestep of $\delta t=10^{-4}$, a branching time of $t_\mathrm{int}=10\delta t$, and an observation time $t=30 t_\mathrm{int}$.

 Each estimate for the CGF at a specific lambda is the mean from 3 runs. They were checked for convergence in walker number and time \cite{hidalgo2017finite,ray2018importance}. For the simulations of ABPs in Fig. 3, we used cloning with guiding forces to accelerate convergence of the estimate \cite{ray2018exact,nemoto2019optimizing}. This was done by adding the non-interacting control force $\bbf{u_{\lambda}}$ to the equations of motion and using cloning with the weight $\Delta W$ (see Eq. 5). The full CGF was then obtained by adding back the non-interacting CGF, $\psi_f(\lambda)$.

In Fig. \ref{fig:S1}, we show the convergence in walkers ($N_w$) for negative $\lambda$. The critical $\lambda$ is close to $-0.0015$. In the limit that the walkers go to infinity the hysteresis seen in the dip around the critical $\lambda$ will disappear. In Figs. \ref{fig:S3}, \ref{fig:S4}, \ref{fig:S5} we show the convergence as a function of walker number for $\lambda=-6.94\times 10^{-4}$, $\lambda=-0.0021$, and $\lambda=-0.0028$ which are before, close to, and after the critical point. The CGF estimate was easily converged for positive $\lambda$ for $1.2\times 10^{4}$ walkers which is consistent with \cite{nemoto2019optimizing}. 

To compute $I_b(s)$, we require a numerical estimate of $\left \langle \Delta W \right \rangle_{\bbf{u}_{\lambda}}$, which was computed for all systems with $N=40$ particles for an observation time of $t=10^{6}\delta t$. The observation time and particle number were increased until convergence of the running average was obtained. The codes used to generate the data in this paper can be found at: \url{https://github.com/kklymko/active_work}

\begin{figure}[h!]
\centering
\includegraphics[scale=0.40]{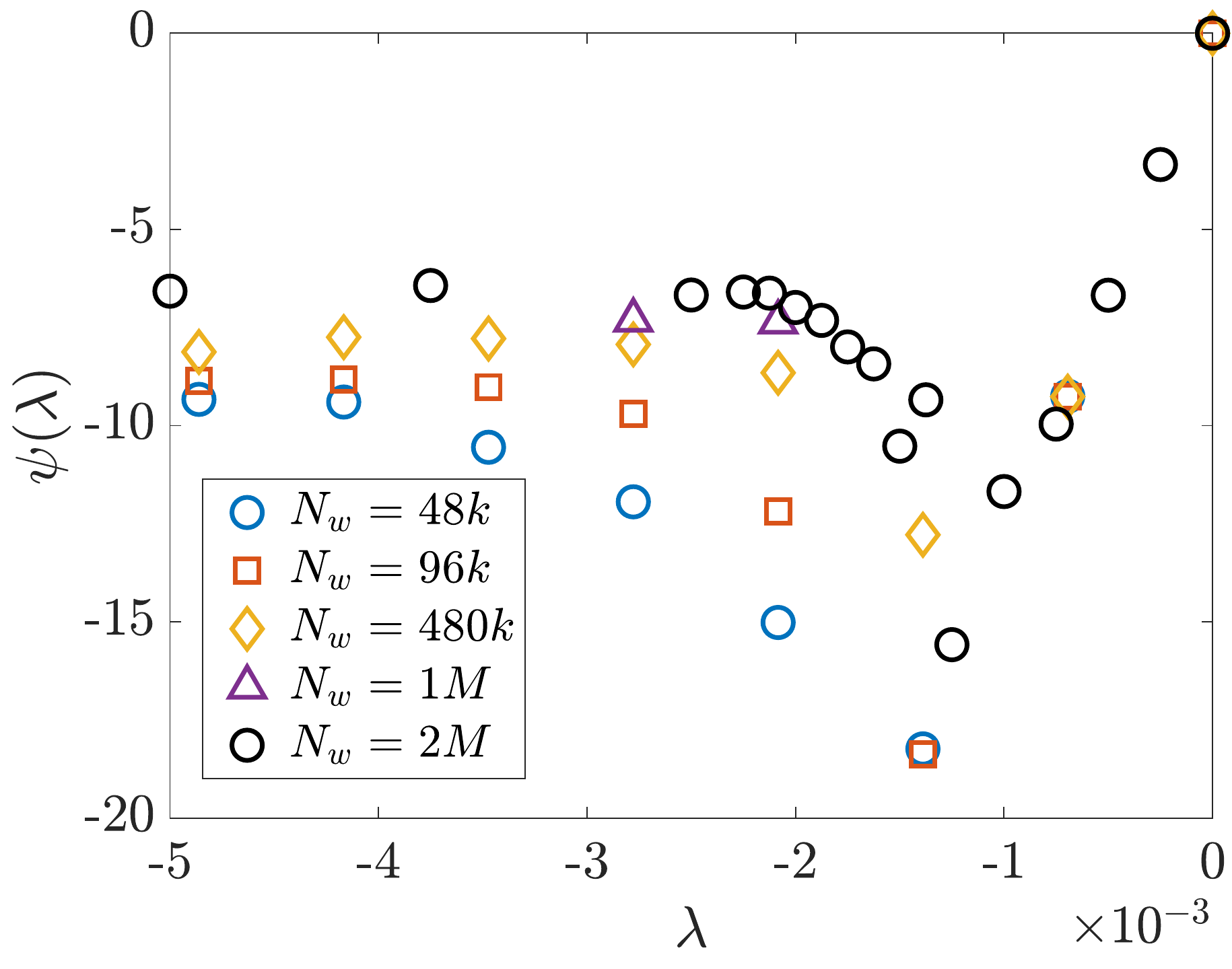}
\caption{
The convergence of the estimate of the CGF for $N=40$, $v=120$ for $t=500 \,\Delta t$ for different numbers of walkers $N_w$.}\label{fig:S1}
\end{figure}
\begin{figure}[h!]
\centering
\includegraphics[scale=0.40]{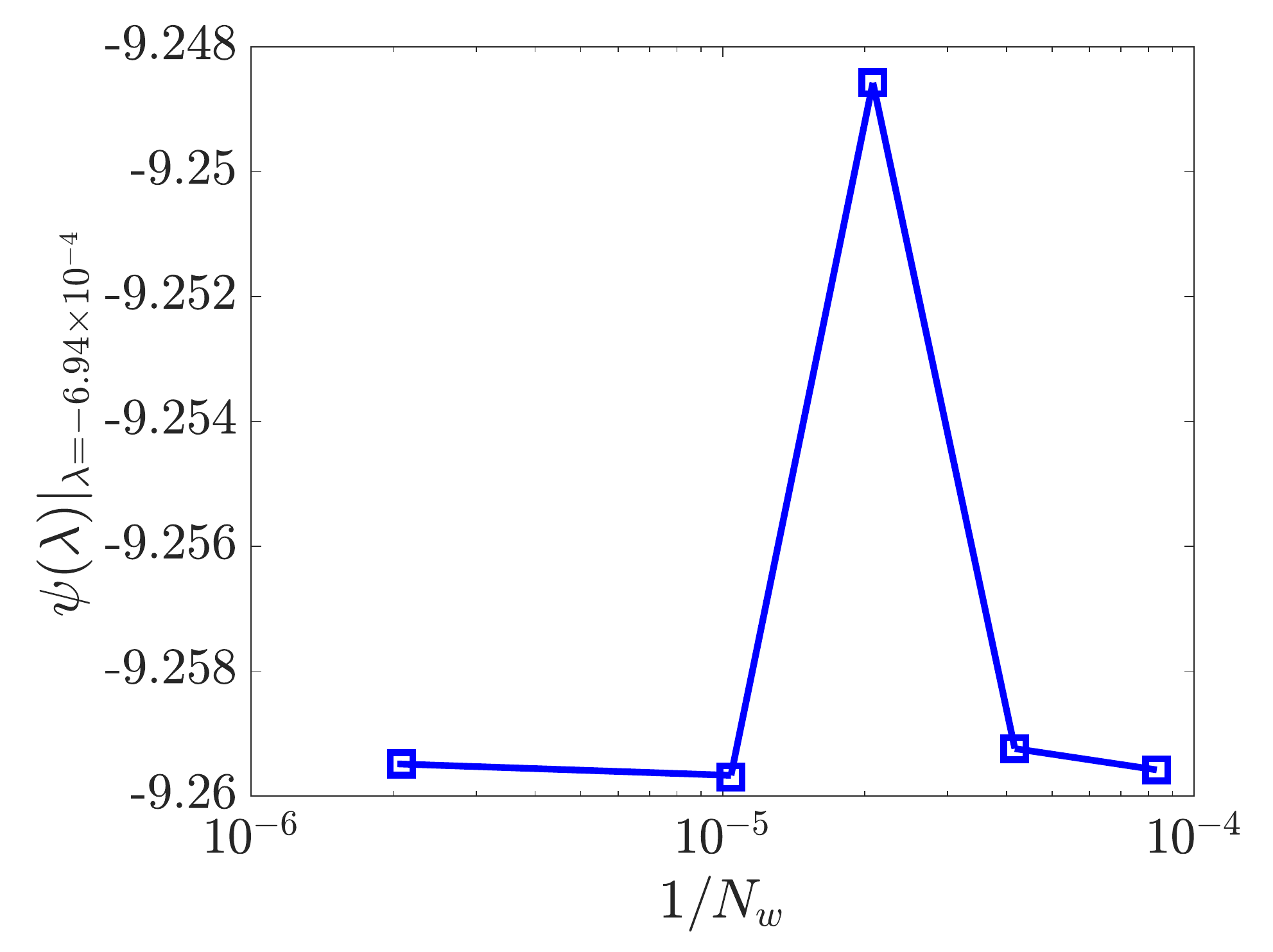}
\caption{
Convergence of the CGF estimate for $v=120$, $N=40$, $t=500\Delta t$, $\lambda=-6.94 \times 10^{-4}$, which is right before  the phase transition. In theory, the CGF estimate is converged at zero on the y-axis.}\label{fig:S3}
\end{figure}

\begin{figure}[h!]
\centering
\includegraphics[scale=0.40]{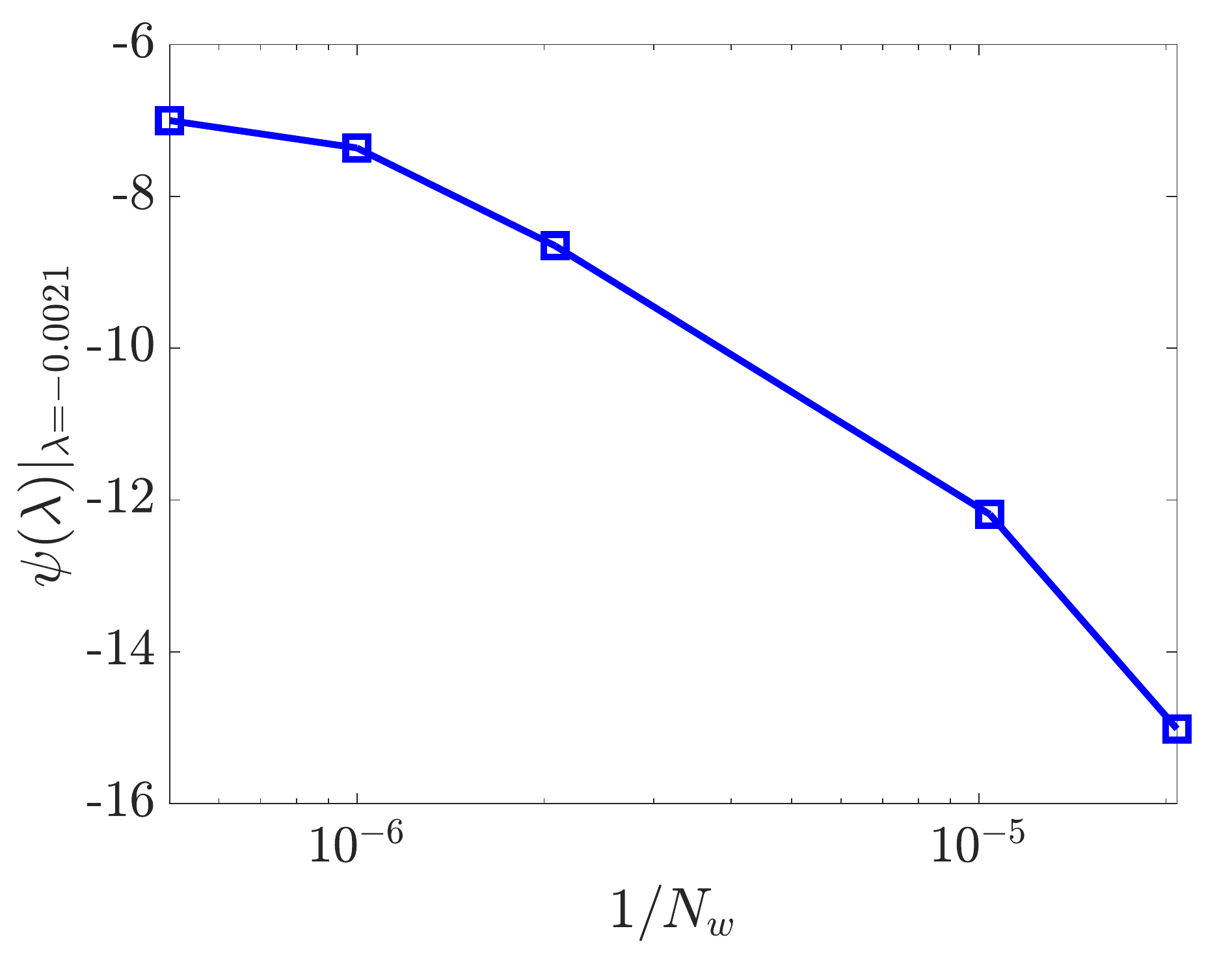}
\caption{
Convergence of the CGF estimate for $v=120$, $N=40$, $t=500\Delta t$, $\lambda=-0.0021$, which is close to $\lambda_c$. In theory, the CGF estimate is converged at zero on the y-axis.}\label{fig:S4}
\end{figure}

\begin{figure}[h!]
\centering
\includegraphics[scale=0.40]{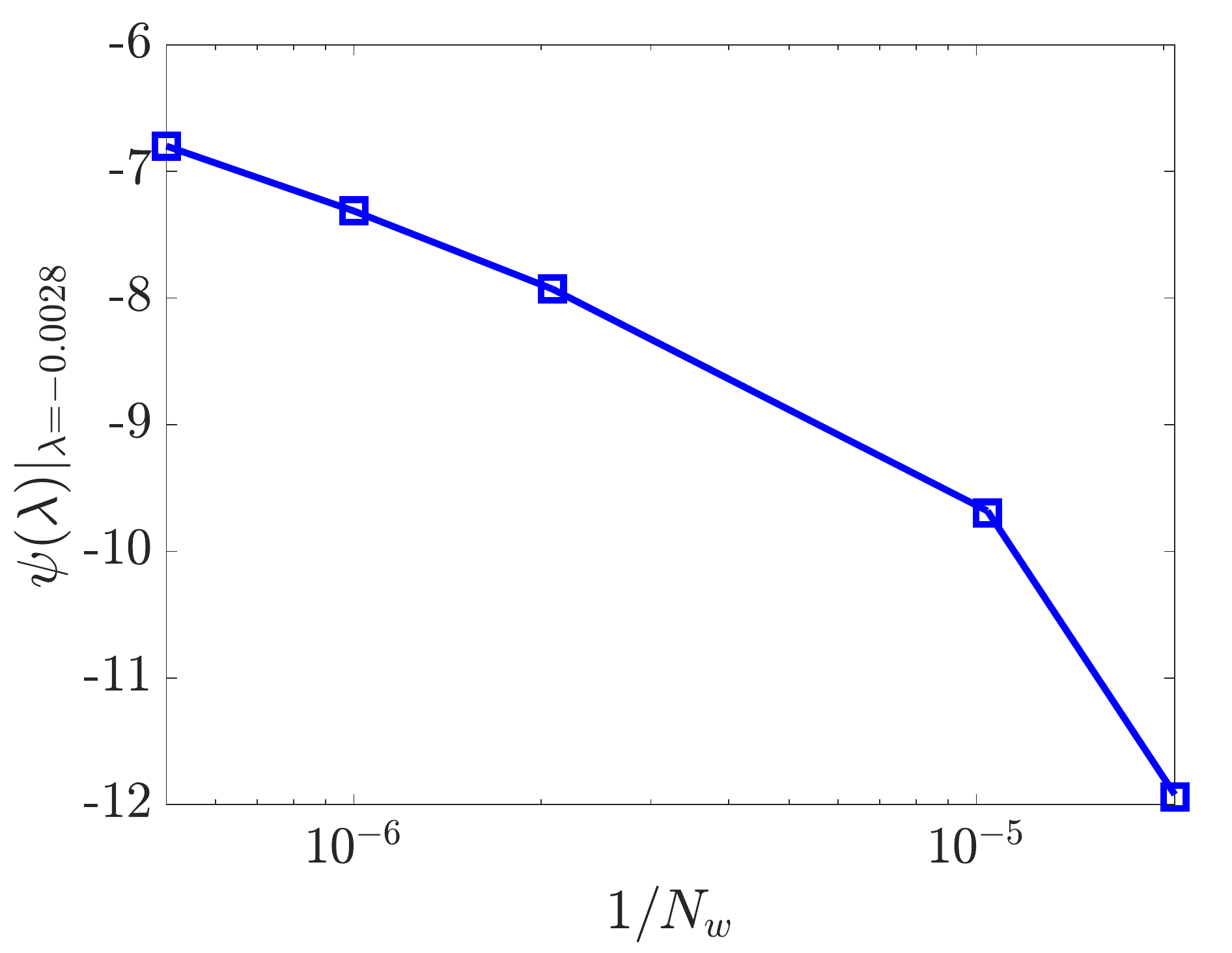}
\caption{
Convergence of the CGF estimate for $v=120$, $N=40$, $t=500\Delta t$, $\lambda=-0.0028$, which is after the phase transition. In theory, the CGF estimate is converged at zero on the y-axis.}\label{fig:S5}
\end{figure}

\subsection{Gaussian fits for Figs. 2 and 3}

In Fig. 2, the  Gaussian fits for small fluctuations for $v=10$ and $v=120$ is $\langle s \rangle_0=v^2(1-0.84\rho)$, $\langle(\delta s)^2\rangle=3v^2$, and $\langle s \rangle_0=v^2(1-0.63\rho)$, and $\langle(\delta s)^2\rangle=12v^2$ with $\rho=0.1$ which are represented by black dotted lines. There is not a clear size dependence for the system sizes studied here and we have found that all three system sizes considered have the same best fit.

In Fig. 3, we fit the dense phase in a  similar way. Although the transition has a system size dependence, once the system is within the phase separated state there is not a clear system size dependence in the variance. The Gaussian fit for the phase separated peak in Figure 3a) and 3b) is given by $\langle s \rangle_0=v^2(1-0.84\rho)$, $\langle(\delta s)^2\rangle=2v^2$ for v=10 with $\rho=0.58$ , and $\langle s \rangle_0=v^2(1-0.84\rho)$, $\langle(\delta s)^2\rangle=2v^2$  for v=120 with $\rho=1.12$. 
Note that the variance for both $v$'s considered for the phase separated system is given by the non-interacting CGF. The averages used in the Maxwell construction and those in Fig. 3c), and d) are slightly different due to the shift in the mean in the thermodynamic limit given by $\langle s\rangle_{\lambda_c}=\langle s\rangle_0+\lambda_c\langle (\delta s)^2\rangle_0$ but the slopes are identical.

\section{General form for the Entropy production}
\label{app:EP}
In order to derive the entropy production for each model, we assume that the self-propulsion is even under time reversal. The difference between choosing the self-propulsion to be even under time reversal is that there is a non-interacting term, as shown in Ref.~\cite{speck2017stochastic}. All of the collective phenomena are thus going to be independent of the convention. 
The distinction between choice of sign is described in more detail in Ref.~\cite{shankar2018hidden}. We note that the convention used in this manuscript is consistent with Ref.~\cite{cagnetta2017large,nemoto2019optimizing,szamel2019stochastic,pietzonka2017entropy}. We also note that the form of the active work is closely related to the swim pressure described in the literature~\cite{takatori2014swim,omar2020microscopic} and can be derived independently from mechanical considerations.

We take the standard definition of the entropy production based on the path probability and its time reversal, 
\begin{equation}
    \Delta S = \ln P[\Gamma]/P[\tilde{\Gamma}] ,
\end{equation}
$P[\Gamma]$ is the probability of observing a path denoted $\Gamma=(r^N(t),b^N(t))$, and $\tilde{\Gamma}=(\tilde{r}^N(t),\tilde{b}^N(t))$ is the time reserved path.  In the time reversed path, we change the signs of functions with explicit time dependence, $\dot{\tilde{\bbf{r}}}_i(-t)=-\dot{\bbf{r}}_i(t)$ and $\dot{\tilde{\bbf{b}}}_i(-t)=-\dot{\bbf{b}}_i(t)$. In the subsequent sections, we write out $P[\Gamma]$ for ABPs and AOUPs and their corresponding entropy production. The ADPs and RTPs can be derived analogously. It is found that all models considered have the same form of $\Delta S$ in the long time limit. 

\subsubsection{ABPs}
The probability of observing a path for a system of ABPs with conservative interactions  in the Stratonovich convention is
\begin{multline}
   P[\Gamma]\propto \exp\Bigg[-\sum_{i=1}^N \int_0^t dt' \ \frac{\bigg(  \bbf{\dot{r}}_i-v\,\bbf{b}_i-\mu\bbf{F}_i \left(\bbf{r}^N \right)\bigg)^2}{4\Dt} \\+\frac{\boldsymbol{\nabla}_{\bbf{r}_i}\cdot\bigg(\mu\bbf{F}_i \left(\bbf{r}^N\right)\bigg)}{2} +\frac{\dot{\bbf{b}}_i^2}{4\Dr} \Bigg] \ ,
   \label{path_ABP}
\end{multline}
where the gradient term in the second line follows from the Stratonovich convention.  After performing the time reversal operation and taking a ratio of path probabilities, the entropy production then becomes
\begin{equation}
    \Delta S=\frac{1}{\Dt}\sum_{i=1}^N \int_0^t dt^{\prime}\, \left [ v\,\bbf{b}_i\circ\dot{\textbf{r}}_i(t^{\prime}) +\dot{\textbf{r}}_i\circ\mu\bbf{F}_i \left(\bbf{r}^N\right)\right ]\ \ ,
    \label{EP}
\end{equation}
which is a sum of two terms. However, since we are using the Stratonovich convention the chain rule is preserved and the term 
\begin{equation}
\sum_{i=1}^{N}    \int_0^t dt' \dot{\textbf{r}}_i\circ \bbf{F}_i \left(\bbf{r}^N\right)=U(\bbf{r^N}(0))-U(\bbf{r^N}(t)) \ ,
\end{equation}
does not grow with time, unlike the first term. In the long time limit it will become negligible, and can be neglected in the entropy production.
\subsubsection{AOUPs}
For AOUPS using the Stratonovich convention, the derivation of the form of the entropy production follows similarly as for the other models. Specifically, the path probability is
\begin{multline}
      P[\Gamma]\propto\exp\Bigg[-\sum_{i=1}^N\int_0^t dt'\frac{\bigg(  \bbf{\dot{r}}_i-v\,\bbf{b}_i-\mu\bbf{F}_i \left(\bbf{r}^N \right)\bigg)^2}{4\Dt}\\+\frac{\boldsymbol{\nabla}_{\bbf{r}_i}\cdot\bigg(\mu\bbf{F}_i \left(\bbf{r}^N\right)\bigg)}{2}\\+\frac{\left(\dot{\bbf{b}}_i+\Dr\bbf{b}_i\right)^2}{2 D_\mathrm{r}}-\frac{\boldsymbol{\nabla}_{\bbf{b}_i}\cdot \Dr\bbf{b}_i}{2}\Bigg] \, ,
      \label{path_AOUP}
\end{multline}
where the additional force on $\bbf{b}_i$ results in the last two terms. After performing the time reversal operation, the entropy production is 
\begin{multline}
    \Delta S =\sum_{i=1}^N\int_{0}^{t}dt^{\prime} \, \bigg(\frac{v\,\bbf{b}_i \circ \dot{\textbf{r}}_i}{\Dt}+\frac{\dot{\textbf{r}}_i\circ\mu\bbf{F}_i \left(\bbf{r}^N\right)}{\Dt}\\+2 \dot{\bbf{b}}_i\circ\bbf{b}_i \bigg)  \ ,
\end{multline}
where the first two terms are analogous to the ABPs. Both the second term and third term do not grow with time, and so in the long time limit the entropy production reduces to
\begin{equation}
    \Delta S=\frac{v}{\Dt}\sum_{i=1}^N \int_0^t dt'   \  \bbf{b}_i\circ\bbf{\dot{r}}_i \ ,
    \label{EP}
\end{equation}
which is equivalent to the form found for the ABPs.

\section{Free particle CGFs and optimal control forces}
\label{app:free}
The free particle CGF is computable from the solution of a generalized eigenvalue equation of the form
\begin{equation}
    L_{\lambda}\nu_{\lambda}=\psi_{f}(\lambda) \nu_{\lambda} \ ,
    \label{eig1}
\end{equation}
where $L_{\lambda}$ is the Lebowitz-Spohn, or tilted, operator and $\nu_{\lambda}$ and $\psi_{f}(\lambda)$ are the maximum eigenvector eigenvalue pair. The tilted operator is derivable from the time evolution of the CGF and the relation to the spectrum of $L_{\lambda}$ and the CGF follows from the long time limit. Generically, for a current-type variable  \cite{lebowitz1999gallavotti,chetrite2015nonequilibrium} the optimal control force that realizes rare entropy production fluctuation is given by 
\begin{equation}
\bbf{u}_{\lambda}=2 \lambda v \bbf{b} +2 \bbf{D} \cdot \nabla \ln \nu_{\lambda} \ ,
\end{equation}
where $\bbf{D}$ is a matrix of diffusion constants in define in space crossed with the self-propulsion vector dimension and $\nabla=\{\nabla_{\bbf{r}},\nabla_{\bbf{b}} \}$. The optimal control force is encoded in the maximum eigenvector associated with $L_{\lambda}$ \cite{chetrite2015nonequilibrium}.

In order to fully solve the eigenspectrum it is necessary to solve the eigenvalue problem for the adjoint tilted operator \cite{touchette2018introduction}
\begin{equation}
    L^{\dagger}_{\lambda}q_{\lambda}=\psi_{f}(\lambda) q_{\lambda} \ .
      \label{eig2}
\end{equation}
since in general $L_{\lambda}$ is not Hermitian. The boundary conditions of the eigevectors must obey a normalization boundary condition $\nu_{\lambda}(\bbf{b})q_{\lambda}(\bbf{b})\rightarrow 0$ as $b\rightarrow \infty $ \cite{touchette2018introduction}. The boundary condition can equivalently be written as 
\begin{equation}
    \int d\bbf{b} \ q_{\lambda}(\bbf{b} )\nu_{\lambda}(\bbf{b})=1,
\end{equation}
and for convenience we impose that
\begin{equation}
    \int d\bbf{b} \ q_{\lambda}(\bbf{b})=1.
\end{equation}
\subsubsection{ABPs}
The tilted generator for the entropy production of an isolated ABP is 
\begin{align}
L_{\lambda} &= v\,\bbf{b}\cdot \bigg[\nabla_{\bbf{r}}+\lambda \frac{ v\,\bbf{b}}{\Dt}\bigg] \\ \nonumber
&\hspace{2 mm}+\Dt \bigg[\nabla_{\bbf{r}}+\lambda \frac{ v\,\bbf{b}}{\Dt}\bigg] \cdot \bigg[\nabla_{\bbf{r}}+\lambda \frac{ v\,\bbf{b}}{\Dt}\bigg]+\bbf{\nabla}_{\bbf{b}}^2\Dr \ ,
\end{align}
which can be solved on a periodic domain by a constant eigenvector, $\nu_\lambda=\mathrm{const}$. This is equivalent to assuming that the stationary state is uniform and isotropic for all $\lambda$. The CGF follows by noting $\bbf{b}\cdot \bbf{b}=1$ and is
\begin{equation}
   \psi_f(\lambda)= \lambda\frac{v^2}{\Dt}+\lambda^2\frac{v^2}{\Dt} \ ,
\end{equation}
and that the control force that realizes the rare dynamics reduces to $\bbf{u}_{\lambda}=2\lambda v\,\bbf{b}$. The corresponding equation of motion is
\begin{equation}
    \bbf{\dot{r}}_i=v\left(1+2\lambda\right)\,\bbf{b}_i+\sqrt{2 \Dt}\bbf{\eta}_i \ ,
    \label{dynamics1}
\end{equation}
where we see explicitly that the control force acts to renormalize the self-propulsion velocity. 

\subsubsection{AOUPs}
 
The tilted generator for the entropy production of an isolated AOUP is 
\begin{multline}
L_{\lambda}=v\,\bbf{b}\cdot \Bigg[\nabla_{\bbf{r}}+\lambda \frac{ v\,\bbf{b}}{\Dt} \Bigg]+\Dt \Bigg[\nabla_{\bbf{r}}+\lambda \frac{ v\,\bbf{b}}{\Dt}\Bigg]\cdot \Bigg[\nabla_{\bbf{r}}+\lambda \frac{ v\,\bbf{b}}{\Dt} \Bigg]\\+\frac{\Dr}{2} \nabla_{\bbf{b}}^2-D_\mathrm{r}\bbf{b}\cdot\nabla_{\bbf{b}} \ ,
\label{AOUPst}
\end{multline}
which contains an additional convective term in $\bbf{b}$ due to the constant restoring force. Assuming the system maintains a uniform and isotropic state at all $\lambda$, such that the eigenvector does not depend on $\bbf{r}$, we can simplify the tilted operator,
\begin{multline}
L_{\lambda}=\lambda\frac{v^2|\bbf{b}|^2}{\Dt}+\lambda^2\frac{v^2|\bbf{b}|^2}{\Dt}+\frac{\Dr}{2}\nabla_{\bbf{b}}^2-D_\mathrm{r}\bbf{b}\cdot\nabla_{\bbf{b}} \ ,
\label{AOUPst1}
\end{multline}
where $b$ is the magnitude of the vector $\bbf{b}$. The domain of $b$ is from $0$ to $\infty$, the eigenvector from equation \eqref{eig1} is
\begin{equation}
   \nu_\lambda(\bbf{b})=\exp\left(\frac{|\bbf{b}|^2\psi_f(\lambda)}{2 D_\mathrm{r}}\right) \, ,
\end{equation}
and its corresponding eigenvalue is
\begin{equation}
\psi_f(\lambda)=D_\mathrm{r}\left(1-\sqrt{1-\frac{2 v^2}{\Dr D_\mathrm{t}}\lambda(1+\lambda)}\right) \ ,
\label{free_AOUP}
\end{equation}
which can be verified by inserting $\nu(\bbf{b})$ back into Eq.~\ref{AOUPst1} 
and noting that since it's in two dimensions it is split up into the $x$ and $y$ dimensions with $|\bbf{b}|^2=\bbf{b}\cdot\bbf{b}=b_x^2+b_y^2$ and $\nabla_{\bbf{b}}^2=\nabla_{b_x}^2+\nabla_{b_y}^2$. The left eigenvector can also be solved to obtain the normalization constant but it is not needed for the control force calculations. 

The optimal control force in the $\bbf{r}$ and $\bbf{b}$ directions, $\bbf{u}_\lambda =  \{ \bbf{u}_\lambda^{\bbf{r}},\bbf{u}_\lambda^{\bbf{b}} \}$ are 
\begin{equation}
\bbf{u}_{\lambda}=\{ 2\lambda v\,\bbf{b},\bbf{b}\psi_f(\lambda) \} \ ,
\end{equation}
which is the result for the control force for non-interacting AOUPs. The biased equations of motion become 
\begin{equation}
    \bbf{\dot{r}}_i=v\left(1+2\lambda\right)\,\bbf{b}_i+\sqrt{2 \Dt}\bbf{\eta}_i \ ,
    \label{dynamics1}
\end{equation}
and
\begin{equation}
    \boldsymbol{\dot{b}}_i=-D_\mathrm{r}\boldsymbol{b}_i\left(1-\psi_f(\lambda)/\Dr\right)+\sqrt{D_\mathrm{r}}\bbf{\xi}_i \ ,
\end{equation}
where the former is identical for ABPs and the latter is specific to AOUPs.

\section{Entropy bounds from Girsanov transformation}
\label{app:bounds}
The CGF for the entropy production can be rewritten as an average over the biased ensemble by preforming a change of measure, or Girsanov transformation, from the original path ensemble with probability $P[\Gamma]$,
\begin{eqnarray}
\label{eq:reweight}
    \psi(\lambda)&=&\frac{1}{t N}\ln \int \mathcal{D}[\Gamma] P[\Gamma] e^{\lambda \Delta S} \nonumber \\
    &=&\frac{1}{t N}\ln \int \mathcal{D}[\Gamma] \frac{P[\Gamma]}{P_{\bbf{u}_\lambda}[\Gamma]}P_{\bbf{u}_\lambda}[\Gamma] e^{\lambda \Delta S} \nonumber \\
    &=&\frac{1}{t N}\ln\left\langle\frac{P[\Gamma]}{P_{\bbf{u}_\lambda}[\Gamma]} e^{\lambda \Delta S} \right\rangle_{\bbf{u}_\lambda} \ ,
\end{eqnarray}
where $P_{\bbf{u}_\lambda}[\Gamma]$ denotes a path ensemble with an additional force $\bbf{u}_\lambda$ added to the original equations of motion, and $\langle \dots \rangle_{\bbf{u}_\lambda}$ denote ensemble average with respect to that measure.  Using Jensen's inequality, we find a general bound within an arbitrary control ensemble \cite{chetrite2015variational},
\begin{equation}
    \psi(\lambda)\ge \frac{1}{t N} \left (\lambda \left\langle \Delta S \right\rangle_{\bbf{u}_\lambda} + \left\langle \ln \frac{P[\Gamma]}{P_{\bbf{u}_\lambda}[\Gamma]}\right \rangle_{\bbf{u}_\lambda} \right ) \ ,
\end{equation}
which need not be tight. However, below we show how in the systems studied by choosing ${\bbf{u}_\lambda}$ to be the optimal control force for the free particle, we can arrive at the tight bound on the entropy production.

\subsubsection{ABPs}

The relative actions with and without the single particle control force for a system of interacting ABPs is
\begin{multline}
\ln \frac{P[\Gamma]}{P_{\bbf{u}_\lambda}[\Gamma]}=\sum_{i=1}^N\int_0^t dt' \,\frac{v^2\lambda(1+\lambda)}{\Dt}-\frac{v}{\Dt} \lambda\bbf{b}_i\circ \dot{\bbf{r}}_i \\ + \beta v{\lambda\bbf{b}_i\cdot\bbf{F}_i \left(\bbf{r}^N \right)} \ ,
\label{ABPweight}
\end{multline}
which employs the identity $\bbf{b}\cdot \bbf{b}=1$. We recognize the first term on the right hand side as $\psi_f(\lambda)$, the second term as the negative of the entropy production, and the final term as $\Delta W$. 
Inserting this relative action into Eq.~\ref{eq:reweight}, we note that the entropy production terms cancel, and we can pull the constants out of the average. The bound can be shown to work analogously for the ADP and RTP models since the added control force does not change the statistics of the orientation, $\bbf{b}$, and only changes the positional degrees of freedom.

\subsubsection{AOUPs}
The relative actions with and without the single particle control force for a system of interacting AOUPs is
\begin{multline}
\ln \frac{P[\Gamma]}{P_{\bbf{u}_\lambda}[\Gamma]}=\\
\int_0^tdt'\left( \frac{v^2 \lambda(1+\lambda)}{\Dt}- \psi_f(1-\psi_f/2\Dr) \right ) b^2\\
+\psi_f(\lambda) -\frac{v}{\Dt}\lambda \bbf{b}_i \circ \bbf{\dot{r}}_i +v\beta{\lambda\bbf{b_i}\cdot\bbf{F}_i \left(\bbf{r}^N \right)}\ ,
     \label{AOUPsre}
\end{multline}
which is more complicated than for the ABPs due to the fluctuating magnitude of the self-propulsion vector. We still can identify the same structure as before, with the free particle CGF, negative of the entropy production, and $\Delta W$, however there is an additional first term in the parenthesis. Inserting the definition of $\psi_f$ from Eq.~\ref{free_AOUP} we find that the term proportional to $v^2 b^2$ is identically 0. This leaves us with the bound for AOUPs.

%

\section{Entropy production from coarse-grained density field}
\label{app:density}
Here we elaborate on our coarse grained theory of the interacting term. Assuming that the important contributions to the interparticle entropy production come from forces that directly oppose self-propulsion, we approximate $\bbf{b}_i \cdot  \bbf{F}(\bbf{r}_{ij}) \approx  -F(r_{ij})$ where $\bbf{F}(\bbf{r}_{ij})$ is the contribution of the $i$th particle's force due to particle $j$ and $\bbf{r}_{ij}$ is the displacement vector between particles $i$ and $j$ with magnitude $r_{ij}$. As presented, under this approximation the fluctuations of $\Delta W$ depend only on the time evolution of the density field. Below we first derive an approximate equation of motion for the density, in the limit of small $k$ and small fluctuations from its mean. Then we describe the approximate calculation of the cumulant generating function and control force.

\subsection{Equation of motion for the density}
We are interested in the density fluctuations with the added control force which changes the self propulsion speed proportional to lambda as $v_{\lambda}=v(1+2\lambda)$. To arrive at an effective equation of motion for the density we first define the instantaneous density field as,
\begin{equation}
    \rho(\bbf{r},t) = \sum_{i=1}^N \delta [\bbf{r}-\bbf{r}_i(t)] \ ,
    \label{fluctuatingrho}
\end{equation}
and corresponding polarization field as
\begin{equation}
    \bbf{P}(\bbf{r},t) = \sum_{i=1}^N \delta [\bbf{r}-\bbf{r}_i(t)]\bbf{b_i}(t) \ ,
\end{equation}
where $\delta$ are Dirac's delta function. In principle, higher order multipoles in the orientation field are needed to completely describe the dynamics, however we neglect quadrupole and higher fields. For the homogeneous states considered, this has been shown to be a good approximation~\cite{solon2015active,bialke2013microscopic}. Following the standard procedures~\cite{dean1996langevin,solon2015pressure} a set of coupled stochastic equation of motion for both fields. For the density field,
\begin{multline}
\frac{\partial \rho(\bbf{r},t)}{\partial t}=-{\nabla_{\bbf{r}}}\left[\mu\rho(\bbf{r},t)\int d\bbf{r}^{\prime}\bbf{F}(\bbf{r}-\bbf{r^{\prime}})\rho(\bbf{r^{\prime}},t)\right . \\\left .+v_{\lambda}\bbf{P}(\bbf{r},t)\right] +D_t\nabla^2_{\bbf{r}}\rho(\bbf{r},t)+\nabla_r\sqrt{2\Delta_{\lambda}}\bbf{\eta}_{\rho}(\boldsymbol{r},t)
\label{densityfull}
\end{multline}
where $\Delta_{\lambda}=\Dt\rho(\bbf{r},t)$ is the mobility and the noise obeys the statistics $\langle \boldsymbol{\eta_{\rho}}(\boldsymbol{r},t)\rangle=0$ and $\langle {\eta^\alpha_{\rho}}(\boldsymbol{r},t){\eta^\beta_{\rho}}(\boldsymbol{r'},t')\rangle=\delta_{\alpha,\beta}\delta(t-t')\delta(\bbf{r}-\bbf{r}')$.
For the polarization field,
\begin{multline}
\frac{\partial \bbf{P}(\bbf{r},t)}{\partial t}=-{\nabla_{\bbf{r}}}\left[\mu \bbf{P}(\bbf{r},t)\int d\bbf{r}^{\prime}F(\bbf{r}-\bbf{r^{\prime}})\rho(\bbf{r^{\prime}},t)\right]\\-{\nabla_{\bbf{r}}} \frac{v_{\lambda}\rho(\bbf{r},t)}{2} +D_t\nabla^2_{\bbf{r}}\bbf{P}(\bbf{r},t)\\ -\Dr \bbf{P}(\bbf{r},t)+\nabla_{\bbf{r}}\sqrt{2\Lambda_P}\bbf{\eta}_{P}(\bbf{r},t)
\end{multline}
where $\eta_P(\bbf{r},t)$ has the same noise statistics as $\boldsymbol{\eta}_{\rho}$ and $\Lambda_P=D_t\bbf{P}(\bbf{r},t)$.

We assume there is a separation of time scales between the density field, which we assume to be slow, and the polarization field, which we assume to relax quickly. Further we assume that on the scale of density fluctuations, the polarization is constant and homogeneous \cite{speck2015dynamical}. These so-called adiabatic assumptions are standard in the treatment of instabilities in the ABP system. Under these assumptions, the polarization is stationary and can be averaged separately from the density and we can neglect its gradient terms. Rearranging the remaining terms, we have an explicit relation between the polarization and density fields,
\begin{equation}
    P(\bbf{r},t)=-\frac{v_{\lambda}}{2\Dr}\nabla_{\bbf{r}} \rho(\bbf{r},t)  \ ,
\end{equation}
which effectively separates the evolution of the two fields. Inserting this into Eq.~\eqref{densityfull} we arrive at a closed equation of motion for the density,
\begin{multline}
\frac{\partial \rho(\bbf{r},t)}{\partial t}=-{\nabla_{\bbf{r}}}\left[\mu\rho(\bbf{r},t)\int d\bbf{r}^{\prime}\bbf{F}(\bbf{r}-\bbf{r^{\prime}})\rho(\bbf{r^{\prime}},t)\right]\\+\mathcal{D_{\lambda}}\nabla^2_{\bbf{r}}\rho(\bbf{r},t)+\nabla_{\bbf{r}}\sqrt{2\Delta_{\lambda}}\bbf{\eta}_{\rho}(\boldsymbol{r},t)
\end{multline}
with $\mathcal{D_{\lambda}}=\Dt+v_{\lambda}^2/2\Dr$ as the effective diffusion constant.

While the equation is closed, it is still nonlinear due to the fluctuating convective term from the interparticle interactions. While more sophisticated expansions exist, for the low densities we consider, we can linearize the evolution equation by simply dropping the second order term in the density, 
\begin{equation}
    \frac{\partial \rho(\bbf{r},t)}{\partial t}={\mathcal{D_{\lambda}}\nabla^2_{\bbf{r}}}\rho(\bbf{r},t)+\nabla_{\bbf{r}}\sqrt{2\Delta_{\lambda}}\eta_{\rho}(\boldsymbol{r},t) \ ,
\end{equation}
which results in a standard fluctuating diffusion equation. Corrections due to interactions can be included phenomenologically by making $\mathcal{D_{\lambda}}$ and $\Delta_{\lambda}$ depend on the mean density. 

Introducing the Fourier transforms, for the density
\begin{equation}
   \hat{\rho}(\bbf{k},t)=\int d\bbf{r} e^{-i\bbf{k}\cdot \bbf{r}}\rho(\bbf{r},t) \ ,
\end{equation}
and the noise,
\begin{equation}
    \boldsymbol{\hat{\eta}}(\bbf{k},t)=\int d\bbf{r} e^{-i\bbf{k} \cdot \bbf{r}} \boldsymbol{\eta}(\bbf{r},t) \ ,
\end{equation}
we can arrive at the equation of motion in Eq. 10.
\vspace{0.2cm}

\subsection{CGF and optimal control force}

The equation of motion for the Fourier transformed density takes the form of a set of uncoupled, complex Ornstein-Uhlenbeck processes for each wavevector. The large deviations of such a system for observables like $\Delta W$ have been considered in detail in Ref.~\onlinecite{dolezal2019large}. The tilted operator for which the CGF of $\Delta W$ is the largest eigenvalue and has the form
\begin{equation}
\mathcal{L}_\lambda = \sum_{k>0}  -k^2\mathcal{D}_\lambda \hat{\rho}_k\nabla_{\hat{\rho}_k} +  k^2  \Delta_\lambda \nabla_{\hat{\rho}_k}^2 -\lambda \frac{\beta v_{\lambda}}{2} \hat{F}(k)\left|\hat{\rho}_k\right|^2 \ ,
\end{equation}
which has to be solved for both the real and imaginary parts of $\hat{\rho}_k$. This can be done following the method of Ref.~\onlinecite{dolezal2019large}. The resulting CGF is
\begin{equation}
\Delta \psi(\lambda) = \frac{1}{N} \sum_{k>0}\Delta \psi_k(\lambda) \ ,
\label{Deltapsi}
\end{equation}
where for each $k$,
\begin{equation}
\Delta \psi_k(\lambda)= k^2 \mathcal{D}_\lambda  \left [1-\sqrt{1+\frac{\beta \lambda v_{\lambda} \hat{F}(k)\Delta_\lambda }{\mathcal{D}_\lambda^2 k^2}} \right ] \ , 
\end{equation}
and the corresponding eigenvector  
\begin{equation}
   \nu_{\lambda}=\prod_{k>0} \exp \left [ \frac{\left|\rho(k,t)\right|^2}{2\Delta_{\lambda}k^2}\Delta \psi_k(\lambda) \right ] \ ,
\end{equation}
factorizes into a product of independent modes, each quadratic in the density.
For a density type variable, the optimal control force is a gradient force, and so can be written as a potential. It is computable following Refs.~\onlinecite{dolezal2019large,angeletti2016diffusions}, 
which in the limit that $\lambda$ approaches zero we recover Eq. 13.


\begin{thebibliography}{96}%
\makeatletter
\providecommand \@ifxundefined [1]{%
 \@ifx{#1\undefined}
}%
\providecommand \@ifnum [1]{%
 \ifnum #1\expandafter \@firstoftwo
 \else \expandafter \@secondoftwo
 \fi
}%
\providecommand \@ifx [1]{%
 \ifx #1\expandafter \@firstoftwo
 \else \expandafter \@secondoftwo
 \fi
}%
\providecommand \natexlab [1]{#1}%
\providecommand \enquote  [1]{``#1''}%
\providecommand \bibnamefont  [1]{#1}%
\providecommand \bibfnamefont [1]{#1}%
\providecommand \citenamefont [1]{#1}%
\providecommand \href@noop [0]{\@secondoftwo}%
\providecommand \href [0]{\begingroup \@sanitize@url \@href}%
\providecommand \@href[1]{\@@startlink{#1}\@@href}%
\providecommand \@@href[1]{\endgroup#1\@@endlink}%
\providecommand \@sanitize@url [0]{\catcode `\\12\catcode `\$12\catcode
  `\&12\catcode `\#12\catcode `\^12\catcode `\_12\catcode `\%12\relax}%
\providecommand \@@startlink[1]{}%
\providecommand \@@endlink[0]{}%
\providecommand \url  [0]{\begingroup\@sanitize@url \@url }%
\providecommand \@url [1]{\endgroup\@href {#1}{\urlprefix }}%
\providecommand \urlprefix  [0]{URL }%
\providecommand \Eprint [0]{\href }%
\providecommand \doibase [0]{http://dx.doi.org/}%
\providecommand \selectlanguage [0]{\@gobble}%
\providecommand \bibinfo  [0]{\@secondoftwo}%
\providecommand \bibfield  [0]{\@secondoftwo}%
\providecommand \translation [1]{[#1]}%
\providecommand \BibitemOpen [0]{}%
\providecommand \bibitemStop [0]{}%
\providecommand \bibitemNoStop [0]{.\EOS\space}%
\providecommand \EOS [0]{\spacefactor3000\relax}%
\providecommand \BibitemShut  [1]{\csname bibitem#1\endcsname}%
\let\auto@bib@innerbib\@empty
\bibitem [{\citenamefont {Nardini}\ \emph {et~al.}(2017)\citenamefont
  {Nardini}, \citenamefont {Fodor}, \citenamefont {Tjhung}, \citenamefont
  {Van~Wijland}, \citenamefont {Tailleur},\ and\ \citenamefont
  {Cates}}]{nardini2017entropy}%
  \BibitemOpen
  \bibfield  {author} {\bibinfo {author} {\bibfnamefont {C.}~\bibnamefont
  {Nardini}}, \bibinfo {author} {\bibfnamefont {{\'E}.}~\bibnamefont {Fodor}},
  \bibinfo {author} {\bibfnamefont {E.}~\bibnamefont {Tjhung}}, \bibinfo
  {author} {\bibfnamefont {F.}~\bibnamefont {Van~Wijland}}, \bibinfo {author}
  {\bibfnamefont {J.}~\bibnamefont {Tailleur}}, \ and\ \bibinfo {author}
  {\bibfnamefont {M.~E.}\ \bibnamefont {Cates}},\ }\href@noop {} {\bibfield
  {journal} {\bibinfo  {journal} {Physical Review X}\ }\textbf {\bibinfo
  {volume} {7}},\ \bibinfo {pages} {021007} (\bibinfo {year}
  {2017})}\BibitemShut {NoStop}%
\bibitem [{\citenamefont {Fodor}\ \emph {et~al.}(2020)\citenamefont {Fodor},
  \citenamefont {Nemoto},\ and\ \citenamefont
  {Vaikuntanathan}}]{fodor2020dissipation}%
  \BibitemOpen
  \bibfield  {author} {\bibinfo {author} {\bibfnamefont {{\'E}.}~\bibnamefont
  {Fodor}}, \bibinfo {author} {\bibfnamefont {T.}~\bibnamefont {Nemoto}}, \
  and\ \bibinfo {author} {\bibfnamefont {S.}~\bibnamefont {Vaikuntanathan}},\
  }\href@noop {} {\bibfield  {journal} {\bibinfo  {journal} {New Journal of
  Physics}\ }\textbf {\bibinfo {volume} {22}},\ \bibinfo {pages} {013052}
  (\bibinfo {year} {2020})}\BibitemShut {NoStop}%
\bibitem [{\citenamefont {del Junco}\ \emph {et~al.}(2018)\citenamefont {del
  Junco}, \citenamefont {Tociu},\ and\ \citenamefont
  {Vaikuntanathan}}]{del2018energy}%
  \BibitemOpen
  \bibfield  {author} {\bibinfo {author} {\bibfnamefont {C.}~\bibnamefont {del
  Junco}}, \bibinfo {author} {\bibfnamefont {L.}~\bibnamefont {Tociu}}, \ and\
  \bibinfo {author} {\bibfnamefont {S.}~\bibnamefont {Vaikuntanathan}},\
  }\href@noop {} {\bibfield  {journal} {\bibinfo  {journal} {Proceedings of the
  National Academy of Sciences}\ }\textbf {\bibinfo {volume} {115}},\ \bibinfo
  {pages} {3569} (\bibinfo {year} {2018})}\BibitemShut {NoStop}%
\bibitem [{\citenamefont {Gaspard}\ and\ \citenamefont
  {Kapral}(2018)}]{gaspard2018fluctuating}%
  \BibitemOpen
  \bibfield  {author} {\bibinfo {author} {\bibfnamefont {P.}~\bibnamefont
  {Gaspard}}\ and\ \bibinfo {author} {\bibfnamefont {R.}~\bibnamefont
  {Kapral}},\ }\href@noop {} {\bibfield  {journal} {\bibinfo  {journal} {The
  Journal of chemical physics}\ }\textbf {\bibinfo {volume} {148}},\ \bibinfo
  {pages} {134104} (\bibinfo {year} {2018})}\BibitemShut {NoStop}%
\bibitem [{\citenamefont {Dasbiswas}\ \emph {et~al.}(2018)\citenamefont
  {Dasbiswas}, \citenamefont {Mandadapu},\ and\ \citenamefont
  {Vaikuntanathan}}]{dasbiswas2018topological}%
  \BibitemOpen
  \bibfield  {author} {\bibinfo {author} {\bibfnamefont {K.}~\bibnamefont
  {Dasbiswas}}, \bibinfo {author} {\bibfnamefont {K.~K.}\ \bibnamefont
  {Mandadapu}}, \ and\ \bibinfo {author} {\bibfnamefont {S.}~\bibnamefont
  {Vaikuntanathan}},\ }\href@noop {} {\bibfield  {journal} {\bibinfo  {journal}
  {Proceedings of the National Academy of Sciences}\ }\textbf {\bibinfo
  {volume} {115}},\ \bibinfo {pages} {E9031} (\bibinfo {year}
  {2018})}\BibitemShut {NoStop}%
\bibitem [{\citenamefont {Soni}\ \emph {et~al.}(2019)\citenamefont {Soni},
  \citenamefont {Bililign}, \citenamefont {Magkiriadou}, \citenamefont
  {Sacanna}, \citenamefont {Bartolo}, \citenamefont {Shelley},\ and\
  \citenamefont {Irvine}}]{soni2019odd}%
  \BibitemOpen
  \bibfield  {author} {\bibinfo {author} {\bibfnamefont {V.}~\bibnamefont
  {Soni}}, \bibinfo {author} {\bibfnamefont {E.~S.}\ \bibnamefont {Bililign}},
  \bibinfo {author} {\bibfnamefont {S.}~\bibnamefont {Magkiriadou}}, \bibinfo
  {author} {\bibfnamefont {S.}~\bibnamefont {Sacanna}}, \bibinfo {author}
  {\bibfnamefont {D.}~\bibnamefont {Bartolo}}, \bibinfo {author} {\bibfnamefont
  {M.~J.}\ \bibnamefont {Shelley}}, \ and\ \bibinfo {author} {\bibfnamefont
  {W.~T.}\ \bibnamefont {Irvine}},\ }\href@noop {} {\bibfield  {journal}
  {\bibinfo  {journal} {Nature Physics}\ }\textbf {\bibinfo {volume} {15}},\
  \bibinfo {pages} {1188} (\bibinfo {year} {2019})}\BibitemShut {NoStop}%
\bibitem [{\citenamefont {Liao}\ \emph {et~al.}(2020)\citenamefont {Liao},
  \citenamefont {Irvine},\ and\ \citenamefont
  {Vaikuntanathan}}]{liao2020rectification}%
  \BibitemOpen
  \bibfield  {author} {\bibinfo {author} {\bibfnamefont {Z.}~\bibnamefont
  {Liao}}, \bibinfo {author} {\bibfnamefont {W.~T.}\ \bibnamefont {Irvine}}, \
  and\ \bibinfo {author} {\bibfnamefont {S.}~\bibnamefont {Vaikuntanathan}},\
  }\href@noop {} {\bibfield  {journal} {\bibinfo  {journal} {Physical Review
  X}\ }\textbf {\bibinfo {volume} {10}},\ \bibinfo {pages} {021036} (\bibinfo
  {year} {2020})}\BibitemShut {NoStop}%
\bibitem [{\citenamefont {Caballero}\ and\ \citenamefont
  {Cates}(2020)}]{PhysRevLett.124.240604}%
  \BibitemOpen
  \bibfield  {author} {\bibinfo {author} {\bibfnamefont {F.}~\bibnamefont
  {Caballero}}\ and\ \bibinfo {author} {\bibfnamefont {M.~E.}\ \bibnamefont
  {Cates}},\ }\href {\doibase 10.1103/PhysRevLett.124.240604} {\bibfield
  {journal} {\bibinfo  {journal} {Phys. Rev. Lett.}\ }\textbf {\bibinfo
  {volume} {124}},\ \bibinfo {pages} {240604} (\bibinfo {year}
  {2020})}\BibitemShut {NoStop}%
\bibitem [{\citenamefont {Fodor}\ \emph {et~al.}(2016)\citenamefont {Fodor},
  \citenamefont {Nardini}, \citenamefont {Cates}, \citenamefont {Tailleur},
  \citenamefont {Visco},\ and\ \citenamefont {van Wijland}}]{fodor2016far}%
  \BibitemOpen
  \bibfield  {author} {\bibinfo {author} {\bibfnamefont {{\'E}.}~\bibnamefont
  {Fodor}}, \bibinfo {author} {\bibfnamefont {C.}~\bibnamefont {Nardini}},
  \bibinfo {author} {\bibfnamefont {M.~E.}\ \bibnamefont {Cates}}, \bibinfo
  {author} {\bibfnamefont {J.}~\bibnamefont {Tailleur}}, \bibinfo {author}
  {\bibfnamefont {P.}~\bibnamefont {Visco}}, \ and\ \bibinfo {author}
  {\bibfnamefont {F.}~\bibnamefont {van Wijland}},\ }\href@noop {} {\bibfield
  {journal} {\bibinfo  {journal} {Physical review letters}\ }\textbf {\bibinfo
  {volume} {117}},\ \bibinfo {pages} {038103} (\bibinfo {year}
  {2016})}\BibitemShut {NoStop}%
\bibitem [{\citenamefont {Mandal}\ \emph {et~al.}(2017)\citenamefont {Mandal},
  \citenamefont {Klymko},\ and\ \citenamefont {DeWeese}}]{mandal2017entropy}%
  \BibitemOpen
  \bibfield  {author} {\bibinfo {author} {\bibfnamefont {D.}~\bibnamefont
  {Mandal}}, \bibinfo {author} {\bibfnamefont {K.}~\bibnamefont {Klymko}}, \
  and\ \bibinfo {author} {\bibfnamefont {M.~R.}\ \bibnamefont {DeWeese}},\
  }\href@noop {} {\bibfield  {journal} {\bibinfo  {journal} {Physical review
  letters}\ }\textbf {\bibinfo {volume} {119}},\ \bibinfo {pages} {258001}
  (\bibinfo {year} {2017})}\BibitemShut {NoStop}%
\bibitem [{\citenamefont {Metselaar}\ \emph {et~al.}(2019)\citenamefont
  {Metselaar}, \citenamefont {Yeomans},\ and\ \citenamefont
  {Doostmohammadi}}]{metselaar2019topology}%
  \BibitemOpen
  \bibfield  {author} {\bibinfo {author} {\bibfnamefont {L.}~\bibnamefont
  {Metselaar}}, \bibinfo {author} {\bibfnamefont {J.~M.}\ \bibnamefont
  {Yeomans}}, \ and\ \bibinfo {author} {\bibfnamefont {A.}~\bibnamefont
  {Doostmohammadi}},\ }\href@noop {} {\bibfield  {journal} {\bibinfo  {journal}
  {Physical Review Letters}\ }\textbf {\bibinfo {volume} {123}},\ \bibinfo
  {pages} {208001} (\bibinfo {year} {2019})}\BibitemShut {NoStop}%
\bibitem [{\citenamefont {Li}\ and\ \citenamefont {ten
  Wolde}(2019)}]{li2019shape}%
  \BibitemOpen
  \bibfield  {author} {\bibinfo {author} {\bibfnamefont {Y.}~\bibnamefont
  {Li}}\ and\ \bibinfo {author} {\bibfnamefont {P.~R.}\ \bibnamefont {ten
  Wolde}},\ }\href@noop {} {\bibfield  {journal} {\bibinfo  {journal} {Physical
  review letters}\ }\textbf {\bibinfo {volume} {123}},\ \bibinfo {pages}
  {148003} (\bibinfo {year} {2019})}\BibitemShut {NoStop}%
\bibitem [{\citenamefont {Wang}\ \emph {et~al.}(2019)\citenamefont {Wang},
  \citenamefont {Guo}, \citenamefont {Tian},\ and\ \citenamefont
  {Chen}}]{wang2019shape}%
  \BibitemOpen
  \bibfield  {author} {\bibinfo {author} {\bibfnamefont {C.}~\bibnamefont
  {Wang}}, \bibinfo {author} {\bibfnamefont {Y.-k.}\ \bibnamefont {Guo}},
  \bibinfo {author} {\bibfnamefont {W.-d.}\ \bibnamefont {Tian}}, \ and\
  \bibinfo {author} {\bibfnamefont {K.}~\bibnamefont {Chen}},\ }\href@noop {}
  {\bibfield  {journal} {\bibinfo  {journal} {The Journal of chemical physics}\
  }\textbf {\bibinfo {volume} {150}},\ \bibinfo {pages} {044907} (\bibinfo
  {year} {2019})}\BibitemShut {NoStop}%
\bibitem [{\citenamefont {Morris}\ and\ \citenamefont
  {Rao}(2019)}]{morris2019active}%
  \BibitemOpen
  \bibfield  {author} {\bibinfo {author} {\bibfnamefont {R.~G.}\ \bibnamefont
  {Morris}}\ and\ \bibinfo {author} {\bibfnamefont {M.}~\bibnamefont {Rao}},\
  }\href@noop {} {\bibfield  {journal} {\bibinfo  {journal} {Physical Review
  E}\ }\textbf {\bibinfo {volume} {100}},\ \bibinfo {pages} {022413} (\bibinfo
  {year} {2019})}\BibitemShut {NoStop}%
\bibitem [{\citenamefont {Bain}\ and\ \citenamefont
  {Bartolo}(2019)}]{bain2019dynamic}%
  \BibitemOpen
  \bibfield  {author} {\bibinfo {author} {\bibfnamefont {N.}~\bibnamefont
  {Bain}}\ and\ \bibinfo {author} {\bibfnamefont {D.}~\bibnamefont {Bartolo}},\
  }\href@noop {} {\bibfield  {journal} {\bibinfo  {journal} {Science}\ }\textbf
  {\bibinfo {volume} {363}},\ \bibinfo {pages} {46} (\bibinfo {year}
  {2019})}\BibitemShut {NoStop}%
\bibitem [{\citenamefont {Morin}\ and\ \citenamefont
  {Bartolo}(2018)}]{morin2018flowing}%
  \BibitemOpen
  \bibfield  {author} {\bibinfo {author} {\bibfnamefont {A.}~\bibnamefont
  {Morin}}\ and\ \bibinfo {author} {\bibfnamefont {D.}~\bibnamefont
  {Bartolo}},\ }\href@noop {} {\bibfield  {journal} {\bibinfo  {journal}
  {Physical Review X}\ }\textbf {\bibinfo {volume} {8}},\ \bibinfo {pages}
  {021037} (\bibinfo {year} {2018})}\BibitemShut {NoStop}%
\bibitem [{\citenamefont {Bricard}\ \emph {et~al.}(2013)\citenamefont
  {Bricard}, \citenamefont {Caussin}, \citenamefont {Desreumaux}, \citenamefont
  {Dauchot},\ and\ \citenamefont {Bartolo}}]{bricard2013emergence}%
  \BibitemOpen
  \bibfield  {author} {\bibinfo {author} {\bibfnamefont {A.}~\bibnamefont
  {Bricard}}, \bibinfo {author} {\bibfnamefont {J.-B.}\ \bibnamefont
  {Caussin}}, \bibinfo {author} {\bibfnamefont {N.}~\bibnamefont {Desreumaux}},
  \bibinfo {author} {\bibfnamefont {O.}~\bibnamefont {Dauchot}}, \ and\
  \bibinfo {author} {\bibfnamefont {D.}~\bibnamefont {Bartolo}},\ }\href@noop
  {} {\bibfield  {journal} {\bibinfo  {journal} {Nature}\ }\textbf {\bibinfo
  {volume} {503}},\ \bibinfo {pages} {95} (\bibinfo {year} {2013})}\BibitemShut
  {NoStop}%
\bibitem [{\citenamefont {Mietke}\ \emph {et~al.}(2019)\citenamefont {Mietke},
  \citenamefont {Jemseena}, \citenamefont {Kumar}, \citenamefont {Sbalzarini},\
  and\ \citenamefont {J{\"u}licher}}]{mietke2019minimal}%
  \BibitemOpen
  \bibfield  {author} {\bibinfo {author} {\bibfnamefont {A.}~\bibnamefont
  {Mietke}}, \bibinfo {author} {\bibfnamefont {V.}~\bibnamefont {Jemseena}},
  \bibinfo {author} {\bibfnamefont {K.~V.}\ \bibnamefont {Kumar}}, \bibinfo
  {author} {\bibfnamefont {I.~F.}\ \bibnamefont {Sbalzarini}}, \ and\ \bibinfo
  {author} {\bibfnamefont {F.}~\bibnamefont {J{\"u}licher}},\ }\href@noop {}
  {\bibfield  {journal} {\bibinfo  {journal} {Physical review letters}\
  }\textbf {\bibinfo {volume} {123}},\ \bibinfo {pages} {188101} (\bibinfo
  {year} {2019})}\BibitemShut {NoStop}%
\bibitem [{\citenamefont {Morozov}(2017)}]{morozov2017chaos}%
  \BibitemOpen
  \bibfield  {author} {\bibinfo {author} {\bibfnamefont {A.}~\bibnamefont
  {Morozov}},\ }\href@noop {} {\bibfield  {journal} {\bibinfo  {journal}
  {Science}\ }\textbf {\bibinfo {volume} {355}},\ \bibinfo {pages} {1262}
  (\bibinfo {year} {2017})}\BibitemShut {NoStop}%
\bibitem [{\citenamefont {Souslov}\ \emph {et~al.}(2017)\citenamefont
  {Souslov}, \citenamefont {Van~Zuiden}, \citenamefont {Bartolo},\ and\
  \citenamefont {Vitelli}}]{souslov2017topological}%
  \BibitemOpen
  \bibfield  {author} {\bibinfo {author} {\bibfnamefont {A.}~\bibnamefont
  {Souslov}}, \bibinfo {author} {\bibfnamefont {B.~C.}\ \bibnamefont
  {Van~Zuiden}}, \bibinfo {author} {\bibfnamefont {D.}~\bibnamefont {Bartolo}},
  \ and\ \bibinfo {author} {\bibfnamefont {V.}~\bibnamefont {Vitelli}},\
  }\href@noop {} {\bibfield  {journal} {\bibinfo  {journal} {Nature Physics}\
  }\textbf {\bibinfo {volume} {13}},\ \bibinfo {pages} {1091} (\bibinfo {year}
  {2017})}\BibitemShut {NoStop}%
\bibitem [{\citenamefont {Ropp}\ \emph {et~al.}(2018)\citenamefont {Ropp},
  \citenamefont {Bachelard}, \citenamefont {Barth}, \citenamefont {Wang},\ and\
  \citenamefont {Zhang}}]{ropp2018dissipative}%
  \BibitemOpen
  \bibfield  {author} {\bibinfo {author} {\bibfnamefont {C.}~\bibnamefont
  {Ropp}}, \bibinfo {author} {\bibfnamefont {N.}~\bibnamefont {Bachelard}},
  \bibinfo {author} {\bibfnamefont {D.}~\bibnamefont {Barth}}, \bibinfo
  {author} {\bibfnamefont {Y.}~\bibnamefont {Wang}}, \ and\ \bibinfo {author}
  {\bibfnamefont {X.}~\bibnamefont {Zhang}},\ }\href@noop {} {\bibfield
  {journal} {\bibinfo  {journal} {Nature Photonics}\ }\textbf {\bibinfo
  {volume} {12}},\ \bibinfo {pages} {739} (\bibinfo {year} {2018})}\BibitemShut
  {NoStop}%
\bibitem [{\citenamefont {Takatori}\ and\ \citenamefont
  {Sahu}(2020)}]{takatori2020active}%
  \BibitemOpen
  \bibfield  {author} {\bibinfo {author} {\bibfnamefont {S.~C.}\ \bibnamefont
  {Takatori}}\ and\ \bibinfo {author} {\bibfnamefont {A.}~\bibnamefont
  {Sahu}},\ }\href@noop {} {\bibfield  {journal} {\bibinfo  {journal} {Physical
  Review Letters}\ }\textbf {\bibinfo {volume} {124}},\ \bibinfo {pages}
  {158102} (\bibinfo {year} {2020})}\BibitemShut {NoStop}%
\bibitem [{\citenamefont {Chaudhuri}(2014)}]{chaudhuri2014active}%
  \BibitemOpen
  \bibfield  {author} {\bibinfo {author} {\bibfnamefont {D.}~\bibnamefont
  {Chaudhuri}},\ }\href@noop {} {\bibfield  {journal} {\bibinfo  {journal}
  {Physical Review E}\ }\textbf {\bibinfo {volume} {90}},\ \bibinfo {pages}
  {022131} (\bibinfo {year} {2014})}\BibitemShut {NoStop}%
\bibitem [{\citenamefont {Pietzonka}\ \emph {et~al.}(2019)\citenamefont
  {Pietzonka}, \citenamefont {Fodor}, \citenamefont {Lohrmann}, \citenamefont
  {Cates},\ and\ \citenamefont {Seifert}}]{PhysRevX.9.041032}%
  \BibitemOpen
  \bibfield  {author} {\bibinfo {author} {\bibfnamefont {P.}~\bibnamefont
  {Pietzonka}}, \bibinfo {author} {\bibfnamefont {E.}~\bibnamefont {Fodor}},
  \bibinfo {author} {\bibfnamefont {C.}~\bibnamefont {Lohrmann}}, \bibinfo
  {author} {\bibfnamefont {M.~E.}\ \bibnamefont {Cates}}, \ and\ \bibinfo
  {author} {\bibfnamefont {U.}~\bibnamefont {Seifert}},\ }\href {\doibase
  10.1103/PhysRevX.9.041032} {\bibfield  {journal} {\bibinfo  {journal} {Phys.
  Rev. X}\ }\textbf {\bibinfo {volume} {9}},\ \bibinfo {pages} {041032}
  (\bibinfo {year} {2019})}\BibitemShut {NoStop}%
\bibitem [{\citenamefont {Krishnamurthy}\ \emph {et~al.}(2016)\citenamefont
  {Krishnamurthy}, \citenamefont {Ghosh}, \citenamefont {Chatterji},
  \citenamefont {Ganapathy},\ and\ \citenamefont
  {Sood}}]{krishnamurthy2016micrometre}%
  \BibitemOpen
  \bibfield  {author} {\bibinfo {author} {\bibfnamefont {S.}~\bibnamefont
  {Krishnamurthy}}, \bibinfo {author} {\bibfnamefont {S.}~\bibnamefont
  {Ghosh}}, \bibinfo {author} {\bibfnamefont {D.}~\bibnamefont {Chatterji}},
  \bibinfo {author} {\bibfnamefont {R.}~\bibnamefont {Ganapathy}}, \ and\
  \bibinfo {author} {\bibfnamefont {A.}~\bibnamefont {Sood}},\ }\href@noop {}
  {\bibfield  {journal} {\bibinfo  {journal} {Nature Physics}\ }\textbf
  {\bibinfo {volume} {12}},\ \bibinfo {pages} {1134} (\bibinfo {year}
  {2016})}\BibitemShut {NoStop}%
\bibitem [{\citenamefont {Zakine}\ \emph {et~al.}(2017)\citenamefont {Zakine},
  \citenamefont {Solon}, \citenamefont {Gingrich},\ and\ \citenamefont {van
  Wijland}}]{zakine2017stochastic}%
  \BibitemOpen
  \bibfield  {author} {\bibinfo {author} {\bibfnamefont {R.}~\bibnamefont
  {Zakine}}, \bibinfo {author} {\bibfnamefont {A.}~\bibnamefont {Solon}},
  \bibinfo {author} {\bibfnamefont {T.}~\bibnamefont {Gingrich}}, \ and\
  \bibinfo {author} {\bibfnamefont {F.}~\bibnamefont {van Wijland}},\
  }\href@noop {} {\bibfield  {journal} {\bibinfo  {journal} {Entropy}\ }\textbf
  {\bibinfo {volume} {19}},\ \bibinfo {pages} {193} (\bibinfo {year}
  {2017})}\BibitemShut {NoStop}%
\bibitem [{\citenamefont {Martin}\ \emph {et~al.}(2018)\citenamefont {Martin},
  \citenamefont {Nardini}, \citenamefont {Cates},\ and\ \citenamefont
  {Fodor}}]{martin2018extracting}%
  \BibitemOpen
  \bibfield  {author} {\bibinfo {author} {\bibfnamefont {D.}~\bibnamefont
  {Martin}}, \bibinfo {author} {\bibfnamefont {C.}~\bibnamefont {Nardini}},
  \bibinfo {author} {\bibfnamefont {M.~E.}\ \bibnamefont {Cates}}, \ and\
  \bibinfo {author} {\bibfnamefont {{\'E}.}~\bibnamefont {Fodor}},\ }\href@noop
  {} {\bibfield  {journal} {\bibinfo  {journal} {EPL (Europhysics Letters)}\
  }\textbf {\bibinfo {volume} {121}},\ \bibinfo {pages} {60005} (\bibinfo
  {year} {2018})}\BibitemShut {NoStop}%
\bibitem [{\citenamefont {Saha}\ \emph {et~al.}(2018)\citenamefont {Saha},
  \citenamefont {Marathe}, \citenamefont {Pal},\ and\ \citenamefont
  {Jayannavar}}]{saha2018stochastic}%
  \BibitemOpen
  \bibfield  {author} {\bibinfo {author} {\bibfnamefont {A.}~\bibnamefont
  {Saha}}, \bibinfo {author} {\bibfnamefont {R.}~\bibnamefont {Marathe}},
  \bibinfo {author} {\bibfnamefont {P.}~\bibnamefont {Pal}}, \ and\ \bibinfo
  {author} {\bibfnamefont {A.}~\bibnamefont {Jayannavar}},\ }\href@noop {}
  {\bibfield  {journal} {\bibinfo  {journal} {Journal of Statistical Mechanics:
  Theory and Experiment}\ }\textbf {\bibinfo {volume} {2018}},\ \bibinfo
  {pages} {113203} (\bibinfo {year} {2018})}\BibitemShut {NoStop}%
\bibitem [{\citenamefont {Ekeh}\ \emph {et~al.}(2020)\citenamefont {Ekeh},
  \citenamefont {Cates},\ and\ \citenamefont {Fodor}}]{PhysRevE.102.010101}%
  \BibitemOpen
  \bibfield  {author} {\bibinfo {author} {\bibfnamefont {T.}~\bibnamefont
  {Ekeh}}, \bibinfo {author} {\bibfnamefont {M.~E.}\ \bibnamefont {Cates}}, \
  and\ \bibinfo {author} {\bibfnamefont {E.}~\bibnamefont {Fodor}},\ }\href
  {\doibase 10.1103/PhysRevE.102.010101} {\bibfield  {journal} {\bibinfo
  {journal} {Phys. Rev. E}\ }\textbf {\bibinfo {volume} {102}},\ \bibinfo
  {pages} {010101} (\bibinfo {year} {2020})}\BibitemShut {NoStop}%
\bibitem [{\citenamefont {Chaki}\ and\ \citenamefont
  {Chakrabarti}(2018)}]{chaki2018entropy}%
  \BibitemOpen
  \bibfield  {author} {\bibinfo {author} {\bibfnamefont {S.}~\bibnamefont
  {Chaki}}\ and\ \bibinfo {author} {\bibfnamefont {R.}~\bibnamefont
  {Chakrabarti}},\ }\href@noop {} {\bibfield  {journal} {\bibinfo  {journal}
  {Physica A: Statistical Mechanics and its Applications}\ }\textbf {\bibinfo
  {volume} {511}},\ \bibinfo {pages} {302} (\bibinfo {year}
  {2018})}\BibitemShut {NoStop}%
\bibitem [{\citenamefont {Jarzynski}(1997)}]{jarzynski1997nonequilibrium}%
  \BibitemOpen
  \bibfield  {author} {\bibinfo {author} {\bibfnamefont {C.}~\bibnamefont
  {Jarzynski}},\ }\href@noop {} {\bibfield  {journal} {\bibinfo  {journal}
  {Physical Review Letters}\ }\textbf {\bibinfo {volume} {78}},\ \bibinfo
  {pages} {2690} (\bibinfo {year} {1997})}\BibitemShut {NoStop}%
\bibitem [{\citenamefont {Crooks}(1999)}]{crooks1999entropy}%
  \BibitemOpen
  \bibfield  {author} {\bibinfo {author} {\bibfnamefont {G.~E.}\ \bibnamefont
  {Crooks}},\ }\href@noop {} {\bibfield  {journal} {\bibinfo  {journal}
  {Physical Review E}\ }\textbf {\bibinfo {volume} {60}},\ \bibinfo {pages}
  {2721} (\bibinfo {year} {1999})}\BibitemShut {NoStop}%
\bibitem [{\citenamefont {Gallavotti}\ and\ \citenamefont
  {Cohen}(1995)}]{gallavotti1995dynamical}%
  \BibitemOpen
  \bibfield  {author} {\bibinfo {author} {\bibfnamefont {G.}~\bibnamefont
  {Gallavotti}}\ and\ \bibinfo {author} {\bibfnamefont {E.~G.~D.}\ \bibnamefont
  {Cohen}},\ }\href@noop {} {\bibfield  {journal} {\bibinfo  {journal}
  {Physical review letters}\ }\textbf {\bibinfo {volume} {74}},\ \bibinfo
  {pages} {2694} (\bibinfo {year} {1995})}\BibitemShut {NoStop}%
\bibitem [{\citenamefont {Gaspard}\ and\ \citenamefont
  {Kapral}(2017)}]{gaspard2017communication}%
  \BibitemOpen
  \bibfield  {author} {\bibinfo {author} {\bibfnamefont {P.}~\bibnamefont
  {Gaspard}}\ and\ \bibinfo {author} {\bibfnamefont {R.}~\bibnamefont
  {Kapral}},\ }\href@noop {} {\bibfield  {journal} {\bibinfo  {journal} {The
  Journal of chemical physics}\ }\textbf {\bibinfo {volume} {147}},\ \bibinfo
  {pages} {211101} (\bibinfo {year} {2017})}\BibitemShut {NoStop}%
\bibitem [{\citenamefont {Horowitz}\ and\ \citenamefont
  {Gingrich}(2019)}]{horowitz2019thermodynamic}%
  \BibitemOpen
  \bibfield  {author} {\bibinfo {author} {\bibfnamefont {J.~M.}\ \bibnamefont
  {Horowitz}}\ and\ \bibinfo {author} {\bibfnamefont {T.~R.}\ \bibnamefont
  {Gingrich}},\ }\href@noop {} {\bibfield  {journal} {\bibinfo  {journal}
  {Nature Physics}\ ,\ \bibinfo {pages} {1}} (\bibinfo {year}
  {2019})}\BibitemShut {NoStop}%
\bibitem [{\citenamefont {Di~Terlizzi}\ and\ \citenamefont
  {Baiesi}(2018)}]{di2018kinetic}%
  \BibitemOpen
  \bibfield  {author} {\bibinfo {author} {\bibfnamefont {I.}~\bibnamefont
  {Di~Terlizzi}}\ and\ \bibinfo {author} {\bibfnamefont {M.}~\bibnamefont
  {Baiesi}},\ }\href@noop {} {\bibfield  {journal} {\bibinfo  {journal}
  {Journal of Physics A: Mathematical and Theoretical}\ }\textbf {\bibinfo
  {volume} {52}},\ \bibinfo {pages} {02LT03} (\bibinfo {year}
  {2018})}\BibitemShut {NoStop}%
\bibitem [{\citenamefont {Owen}\ \emph {et~al.}(2020)\citenamefont {Owen},
  \citenamefont {Gingrich},\ and\ \citenamefont
  {Horowitz}}]{owen2020universal}%
  \BibitemOpen
  \bibfield  {author} {\bibinfo {author} {\bibfnamefont {J.~A.}\ \bibnamefont
  {Owen}}, \bibinfo {author} {\bibfnamefont {T.~R.}\ \bibnamefont {Gingrich}},
  \ and\ \bibinfo {author} {\bibfnamefont {J.~M.}\ \bibnamefont {Horowitz}},\
  }\href@noop {} {\bibfield  {journal} {\bibinfo  {journal} {Physical Review
  X}\ }\textbf {\bibinfo {volume} {10}},\ \bibinfo {pages} {011066} (\bibinfo
  {year} {2020})}\BibitemShut {NoStop}%
\bibitem [{\citenamefont {Nardini}\ and\ \citenamefont
  {Touchette}(2018)}]{nardini2018process}%
  \BibitemOpen
  \bibfield  {author} {\bibinfo {author} {\bibfnamefont {C.}~\bibnamefont
  {Nardini}}\ and\ \bibinfo {author} {\bibfnamefont {H.}~\bibnamefont
  {Touchette}},\ }\href@noop {} {\bibfield  {journal} {\bibinfo  {journal} {The
  European Physical Journal B}\ }\textbf {\bibinfo {volume} {91}},\ \bibinfo
  {pages} {16} (\bibinfo {year} {2018})}\BibitemShut {NoStop}%
\bibitem [{\citenamefont {Gao}\ and\ \citenamefont
  {Limmer}(2019)}]{gao2019nonlinear}%
  \BibitemOpen
  \bibfield  {author} {\bibinfo {author} {\bibfnamefont {C.~Y.}\ \bibnamefont
  {Gao}}\ and\ \bibinfo {author} {\bibfnamefont {D.~T.}\ \bibnamefont
  {Limmer}},\ }\href@noop {} {\bibfield  {journal} {\bibinfo  {journal} {The
  Journal of chemical physics}\ }\textbf {\bibinfo {volume} {151}},\ \bibinfo
  {pages} {014101} (\bibinfo {year} {2019})}\BibitemShut {NoStop}%
\bibitem [{\citenamefont {Dechant}\ and\ \citenamefont
  {Sasa}(2020)}]{dechant2020fluctuation}%
  \BibitemOpen
  \bibfield  {author} {\bibinfo {author} {\bibfnamefont {A.}~\bibnamefont
  {Dechant}}\ and\ \bibinfo {author} {\bibfnamefont {S.-i.}\ \bibnamefont
  {Sasa}},\ }\href@noop {} {\bibfield  {journal} {\bibinfo  {journal}
  {Proceedings of the National Academy of Sciences}\ }\textbf {\bibinfo
  {volume} {117}},\ \bibinfo {pages} {6430} (\bibinfo {year}
  {2020})}\BibitemShut {NoStop}%
\bibitem [{\citenamefont {Polettini}\ and\ \citenamefont
  {Esposito}(2019)}]{polettini2019effective}%
  \BibitemOpen
  \bibfield  {author} {\bibinfo {author} {\bibfnamefont {M.}~\bibnamefont
  {Polettini}}\ and\ \bibinfo {author} {\bibfnamefont {M.}~\bibnamefont
  {Esposito}},\ }\href@noop {} {\bibfield  {journal} {\bibinfo  {journal}
  {Journal of Statistical Physics}\ }\textbf {\bibinfo {volume} {176}},\
  \bibinfo {pages} {94} (\bibinfo {year} {2019})}\BibitemShut {NoStop}%
\bibitem [{\citenamefont {Barbier}\ and\ \citenamefont
  {Gaspard}(2018)}]{barbier2018microreversibility}%
  \BibitemOpen
  \bibfield  {author} {\bibinfo {author} {\bibfnamefont {M.}~\bibnamefont
  {Barbier}}\ and\ \bibinfo {author} {\bibfnamefont {P.}~\bibnamefont
  {Gaspard}},\ }\href@noop {} {\bibfield  {journal} {\bibinfo  {journal}
  {Journal of Physics A: Mathematical and Theoretical}\ }\textbf {\bibinfo
  {volume} {51}},\ \bibinfo {pages} {355001} (\bibinfo {year}
  {2018})}\BibitemShut {NoStop}%
\bibitem [{\citenamefont {Epstein}\ and\ \citenamefont
  {Mandadapu}(2020)}]{epstein2020time}%
  \BibitemOpen
  \bibfield  {author} {\bibinfo {author} {\bibfnamefont {J.~M.}\ \bibnamefont
  {Epstein}}\ and\ \bibinfo {author} {\bibfnamefont {K.~K.}\ \bibnamefont
  {Mandadapu}},\ }\href@noop {} {\bibfield  {journal} {\bibinfo  {journal}
  {Physical Review E}\ }\textbf {\bibinfo {volume} {101}},\ \bibinfo {pages}
  {052614} (\bibinfo {year} {2020})}\BibitemShut {NoStop}%
\bibitem [{\citenamefont {Hargus}\ \emph {et~al.}(2020)\citenamefont {Hargus},
  \citenamefont {Klymko}, \citenamefont {Epstein},\ and\ \citenamefont
  {Mandadapu}}]{hargus2020time}%
  \BibitemOpen
  \bibfield  {author} {\bibinfo {author} {\bibfnamefont {C.}~\bibnamefont
  {Hargus}}, \bibinfo {author} {\bibfnamefont {K.}~\bibnamefont {Klymko}},
  \bibinfo {author} {\bibfnamefont {J.~M.}\ \bibnamefont {Epstein}}, \ and\
  \bibinfo {author} {\bibfnamefont {K.~K.}\ \bibnamefont {Mandadapu}},\
  }\href@noop {} {\bibfield  {journal} {\bibinfo  {journal} {The Journal of
  Chemical Physics}\ }\textbf {\bibinfo {volume} {152}},\ \bibinfo {pages}
  {201102} (\bibinfo {year} {2020})}\BibitemShut {NoStop}%
\bibitem [{\citenamefont {Wagner}\ \emph {et~al.}(2019)\citenamefont {Wagner},
  \citenamefont {Hagan},\ and\ \citenamefont {Baskaran}}]{wagner2019response}%
  \BibitemOpen
  \bibfield  {author} {\bibinfo {author} {\bibfnamefont {C.~G.}\ \bibnamefont
  {Wagner}}, \bibinfo {author} {\bibfnamefont {M.~F.}\ \bibnamefont {Hagan}}, \
  and\ \bibinfo {author} {\bibfnamefont {A.}~\bibnamefont {Baskaran}},\
  }\href@noop {} {\bibfield  {journal} {\bibinfo  {journal} {Physical Review
  E}\ }\textbf {\bibinfo {volume} {100}},\ \bibinfo {pages} {042610} (\bibinfo
  {year} {2019})}\BibitemShut {NoStop}%
\bibitem [{\citenamefont {Asheichyk}\ \emph {et~al.}(2019)\citenamefont
  {Asheichyk}, \citenamefont {Solon}, \citenamefont {Rohwer},\ and\
  \citenamefont {Kr{\"u}ger}}]{asheichyk2019response}%
  \BibitemOpen
  \bibfield  {author} {\bibinfo {author} {\bibfnamefont {K.}~\bibnamefont
  {Asheichyk}}, \bibinfo {author} {\bibfnamefont {A.~P.}\ \bibnamefont
  {Solon}}, \bibinfo {author} {\bibfnamefont {C.~M.}\ \bibnamefont {Rohwer}}, \
  and\ \bibinfo {author} {\bibfnamefont {M.}~\bibnamefont {Kr{\"u}ger}},\
  }\href@noop {} {\bibfield  {journal} {\bibinfo  {journal} {The Journal of
  chemical physics}\ }\textbf {\bibinfo {volume} {150}},\ \bibinfo {pages}
  {144111} (\bibinfo {year} {2019})}\BibitemShut {NoStop}%
\bibitem [{\citenamefont {Dal~Cengio}\ \emph {et~al.}(2019)\citenamefont
  {Dal~Cengio}, \citenamefont {Levis},\ and\ \citenamefont
  {Pagonabarraga}}]{dal2019linear}%
  \BibitemOpen
  \bibfield  {author} {\bibinfo {author} {\bibfnamefont {S.}~\bibnamefont
  {Dal~Cengio}}, \bibinfo {author} {\bibfnamefont {D.}~\bibnamefont {Levis}}, \
  and\ \bibinfo {author} {\bibfnamefont {I.}~\bibnamefont {Pagonabarraga}},\
  }\href@noop {} {\bibfield  {journal} {\bibinfo  {journal} {Physical Review
  Letters}\ }\textbf {\bibinfo {volume} {123}},\ \bibinfo {pages} {238003}
  (\bibinfo {year} {2019})}\BibitemShut {NoStop}%
\bibitem [{\citenamefont {Caprini}\ \emph {et~al.}(2018)\citenamefont
  {Caprini}, \citenamefont {Marconi},\ and\ \citenamefont
  {Vulpiani}}]{caprini2018linear}%
  \BibitemOpen
  \bibfield  {author} {\bibinfo {author} {\bibfnamefont {L.}~\bibnamefont
  {Caprini}}, \bibinfo {author} {\bibfnamefont {U.~M.~B.}\ \bibnamefont
  {Marconi}}, \ and\ \bibinfo {author} {\bibfnamefont {A.}~\bibnamefont
  {Vulpiani}},\ }\href@noop {} {\bibfield  {journal} {\bibinfo  {journal}
  {Journal of Statistical Mechanics: Theory and Experiment}\ }\textbf {\bibinfo
  {volume} {2018}},\ \bibinfo {pages} {033203} (\bibinfo {year}
  {2018})}\BibitemShut {NoStop}%
\bibitem [{\citenamefont {Merlitz}\ \emph {et~al.}(2018)\citenamefont
  {Merlitz}, \citenamefont {Vuijk}, \citenamefont {Brader}, \citenamefont
  {Sharma},\ and\ \citenamefont {Sommer}}]{merlitz2018linear}%
  \BibitemOpen
  \bibfield  {author} {\bibinfo {author} {\bibfnamefont {H.}~\bibnamefont
  {Merlitz}}, \bibinfo {author} {\bibfnamefont {H.~D.}\ \bibnamefont {Vuijk}},
  \bibinfo {author} {\bibfnamefont {J.}~\bibnamefont {Brader}}, \bibinfo
  {author} {\bibfnamefont {A.}~\bibnamefont {Sharma}}, \ and\ \bibinfo {author}
  {\bibfnamefont {J.-U.}\ \bibnamefont {Sommer}},\ }\href@noop {} {\bibfield
  {journal} {\bibinfo  {journal} {The Journal of Chemical Physics}\ }\textbf
  {\bibinfo {volume} {148}},\ \bibinfo {pages} {194116} (\bibinfo {year}
  {2018})}\BibitemShut {NoStop}%
\bibitem [{\citenamefont {Liao}\ \emph {et~al.}(2019)\citenamefont {Liao},
  \citenamefont {Han}, \citenamefont {Fruchart}, \citenamefont {Vitelli},\ and\
  \citenamefont {Vaikuntanathan}}]{liao2019mechanism}%
  \BibitemOpen
  \bibfield  {author} {\bibinfo {author} {\bibfnamefont {Z.}~\bibnamefont
  {Liao}}, \bibinfo {author} {\bibfnamefont {M.}~\bibnamefont {Han}}, \bibinfo
  {author} {\bibfnamefont {M.}~\bibnamefont {Fruchart}}, \bibinfo {author}
  {\bibfnamefont {V.}~\bibnamefont {Vitelli}}, \ and\ \bibinfo {author}
  {\bibfnamefont {S.}~\bibnamefont {Vaikuntanathan}},\ }\href@noop {}
  {\bibfield  {journal} {\bibinfo  {journal} {The Journal of chemical physics}\
  }\textbf {\bibinfo {volume} {151}},\ \bibinfo {pages} {194108} (\bibinfo
  {year} {2019})}\BibitemShut {NoStop}%
\bibitem [{\citenamefont {Chetrite}\ and\ \citenamefont
  {Touchette}(2015{\natexlab{a}})}]{chetrite2015nonequilibrium}%
  \BibitemOpen
  \bibfield  {author} {\bibinfo {author} {\bibfnamefont {R.}~\bibnamefont
  {Chetrite}}\ and\ \bibinfo {author} {\bibfnamefont {H.}~\bibnamefont
  {Touchette}},\ }in\ \href@noop {} {\emph {\bibinfo {booktitle} {Annales Henri
  Poincar{\'e}}}},\ Vol.~\bibinfo {volume} {16}\ (\bibinfo {organization}
  {Springer},\ \bibinfo {year} {2015})\ pp.\ \bibinfo {pages}
  {2005--2057}\BibitemShut {NoStop}%
\bibitem [{\citenamefont {Barato}\ and\ \citenamefont
  {Seifert}(2015)}]{barato2015thermodynamic}%
  \BibitemOpen
  \bibfield  {author} {\bibinfo {author} {\bibfnamefont {A.~C.}\ \bibnamefont
  {Barato}}\ and\ \bibinfo {author} {\bibfnamefont {U.}~\bibnamefont
  {Seifert}},\ }\href@noop {} {\bibfield  {journal} {\bibinfo  {journal}
  {Physical review letters}\ }\textbf {\bibinfo {volume} {114}},\ \bibinfo
  {pages} {158101} (\bibinfo {year} {2015})}\BibitemShut {NoStop}%
\bibitem [{\citenamefont {Gingrich}\ \emph {et~al.}(2016)\citenamefont
  {Gingrich}, \citenamefont {Horowitz}, \citenamefont {Perunov},\ and\
  \citenamefont {England}}]{PhysRevLett.116.120601}%
  \BibitemOpen
  \bibfield  {author} {\bibinfo {author} {\bibfnamefont {T.~R.}\ \bibnamefont
  {Gingrich}}, \bibinfo {author} {\bibfnamefont {J.~M.}\ \bibnamefont
  {Horowitz}}, \bibinfo {author} {\bibfnamefont {N.}~\bibnamefont {Perunov}}, \
  and\ \bibinfo {author} {\bibfnamefont {J.~L.}\ \bibnamefont {England}},\
  }\href {\doibase 10.1103/PhysRevLett.116.120601} {\bibfield  {journal}
  {\bibinfo  {journal} {Phys. Rev. Lett.}\ }\textbf {\bibinfo {volume} {116}},\
  \bibinfo {pages} {120601} (\bibinfo {year} {2016})}\BibitemShut {NoStop}%
\bibitem [{\citenamefont {Lebowitz}\ and\ \citenamefont
  {Spohn}(1999)}]{lebowitz1999gallavotti}%
  \BibitemOpen
  \bibfield  {author} {\bibinfo {author} {\bibfnamefont {J.}~\bibnamefont
  {Lebowitz}}\ and\ \bibinfo {author} {\bibfnamefont {H.}~\bibnamefont
  {Spohn}},\ }\href@noop {} {\bibfield  {journal} {\bibinfo  {journal} {Journal
  of Statistical Physics}\ }\textbf {\bibinfo {volume} {95}},\ \bibinfo {pages}
  {333} (\bibinfo {year} {1999})}\BibitemShut {NoStop}%
\bibitem [{\citenamefont {Seifert}(2005)}]{seifert2005entropy}%
  \BibitemOpen
  \bibfield  {author} {\bibinfo {author} {\bibfnamefont {U.}~\bibnamefont
  {Seifert}},\ }\href@noop {} {\bibfield  {journal} {\bibinfo  {journal}
  {Physical review letters}\ }\textbf {\bibinfo {volume} {95}},\ \bibinfo
  {pages} {040602} (\bibinfo {year} {2005})}\BibitemShut {NoStop}%
\bibitem [{\citenamefont {Maes}\ and\ \citenamefont
  {Neto{\v{c}}n{\`y}}(2003)}]{maes2003time}%
  \BibitemOpen
  \bibfield  {author} {\bibinfo {author} {\bibfnamefont {C.}~\bibnamefont
  {Maes}}\ and\ \bibinfo {author} {\bibfnamefont {K.}~\bibnamefont
  {Neto{\v{c}}n{\`y}}},\ }\href@noop {} {\bibfield  {journal} {\bibinfo
  {journal} {Journal of statistical physics}\ }\textbf {\bibinfo {volume}
  {110}},\ \bibinfo {pages} {269} (\bibinfo {year} {2003})}\BibitemShut
  {NoStop}%
\bibitem [{\citenamefont {Caprini}\ \emph {et~al.}(2019)\citenamefont
  {Caprini}, \citenamefont {Marconi}, \citenamefont {Puglisi},\ and\
  \citenamefont {Vulpiani}}]{caprini2019entropy}%
  \BibitemOpen
  \bibfield  {author} {\bibinfo {author} {\bibfnamefont {L.}~\bibnamefont
  {Caprini}}, \bibinfo {author} {\bibfnamefont {U.~M.~B.}\ \bibnamefont
  {Marconi}}, \bibinfo {author} {\bibfnamefont {A.}~\bibnamefont {Puglisi}}, \
  and\ \bibinfo {author} {\bibfnamefont {A.}~\bibnamefont {Vulpiani}},\
  }\href@noop {} {\bibfield  {journal} {\bibinfo  {journal} {Journal of
  Statistical Mechanics: Theory and Experiment}\ }\textbf {\bibinfo {volume}
  {2019}},\ \bibinfo {pages} {053203} (\bibinfo {year} {2019})}\BibitemShut
  {NoStop}%
\bibitem [{\citenamefont {Chaki}\ and\ \citenamefont
  {Chakrabarti}(2019)}]{chaki2019effects}%
  \BibitemOpen
  \bibfield  {author} {\bibinfo {author} {\bibfnamefont {S.}~\bibnamefont
  {Chaki}}\ and\ \bibinfo {author} {\bibfnamefont {R.}~\bibnamefont
  {Chakrabarti}},\ }\href@noop {} {\bibfield  {journal} {\bibinfo  {journal}
  {Physica A: Statistical Mechanics and its Applications}\ }\textbf {\bibinfo
  {volume} {530}},\ \bibinfo {pages} {121574} (\bibinfo {year}
  {2019})}\BibitemShut {NoStop}%
\bibitem [{\citenamefont {Cagnetta}\ \emph {et~al.}(2017)\citenamefont
  {Cagnetta}, \citenamefont {Corberi}, \citenamefont {Gonnella},\ and\
  \citenamefont {Suma}}]{cagnetta2017large}%
  \BibitemOpen
  \bibfield  {author} {\bibinfo {author} {\bibfnamefont {F.}~\bibnamefont
  {Cagnetta}}, \bibinfo {author} {\bibfnamefont {F.}~\bibnamefont {Corberi}},
  \bibinfo {author} {\bibfnamefont {G.}~\bibnamefont {Gonnella}}, \ and\
  \bibinfo {author} {\bibfnamefont {A.}~\bibnamefont {Suma}},\ }\href@noop {}
  {\bibfield  {journal} {\bibinfo  {journal} {Physical review letters}\
  }\textbf {\bibinfo {volume} {119}},\ \bibinfo {pages} {158002} (\bibinfo
  {year} {2017})}\BibitemShut {NoStop}%
\bibitem [{\citenamefont {Nemoto}\ \emph {et~al.}(2019)\citenamefont {Nemoto},
  \citenamefont {Fodor}, \citenamefont {Cates}, \citenamefont {Jack},\ and\
  \citenamefont {Tailleur}}]{nemoto2019optimizing}%
  \BibitemOpen
  \bibfield  {author} {\bibinfo {author} {\bibfnamefont {T.}~\bibnamefont
  {Nemoto}}, \bibinfo {author} {\bibfnamefont {{\'E}.}~\bibnamefont {Fodor}},
  \bibinfo {author} {\bibfnamefont {M.~E.}\ \bibnamefont {Cates}}, \bibinfo
  {author} {\bibfnamefont {R.~L.}\ \bibnamefont {Jack}}, \ and\ \bibinfo
  {author} {\bibfnamefont {J.}~\bibnamefont {Tailleur}},\ }\href@noop {}
  {\bibfield  {journal} {\bibinfo  {journal} {Physical Review E}\ }\textbf
  {\bibinfo {volume} {99}},\ \bibinfo {pages} {022605} (\bibinfo {year}
  {2019})}\BibitemShut {NoStop}%
\bibitem [{\citenamefont {Shankar}\ and\ \citenamefont
  {Marchetti}(2018)}]{shankar2018hidden}%
  \BibitemOpen
  \bibfield  {author} {\bibinfo {author} {\bibfnamefont {S.}~\bibnamefont
  {Shankar}}\ and\ \bibinfo {author} {\bibfnamefont {M.~C.}\ \bibnamefont
  {Marchetti}},\ }\href@noop {} {\bibfield  {journal} {\bibinfo  {journal}
  {Physical Review E}\ }\textbf {\bibinfo {volume} {98}},\ \bibinfo {pages}
  {020604} (\bibinfo {year} {2018})}\BibitemShut {NoStop}%
\bibitem [{\citenamefont {Szamel}(2019)}]{szamel2019stochastic}%
  \BibitemOpen
  \bibfield  {author} {\bibinfo {author} {\bibfnamefont {G.}~\bibnamefont
  {Szamel}},\ }\href@noop {} {\bibfield  {journal} {\bibinfo  {journal}
  {Physical Review E}\ }\textbf {\bibinfo {volume} {100}},\ \bibinfo {pages}
  {050603} (\bibinfo {year} {2019})}\BibitemShut {NoStop}%
\bibitem [{\citenamefont {Pietzonka}\ and\ \citenamefont
  {Seifert}(2017)}]{pietzonka2017entropy}%
  \BibitemOpen
  \bibfield  {author} {\bibinfo {author} {\bibfnamefont {P.}~\bibnamefont
  {Pietzonka}}\ and\ \bibinfo {author} {\bibfnamefont {U.}~\bibnamefont
  {Seifert}},\ }\href@noop {} {\bibfield  {journal} {\bibinfo  {journal}
  {Journal of Physics A: Mathematical and Theoretical}\ }\textbf {\bibinfo
  {volume} {51}},\ \bibinfo {pages} {01LT01} (\bibinfo {year}
  {2017})}\BibitemShut {NoStop}%
\bibitem [{\citenamefont {Dabelow}\ \emph {et~al.}(2019)\citenamefont
  {Dabelow}, \citenamefont {Bo},\ and\ \citenamefont
  {Eichhorn}}]{dabelow2019irreversibility}%
  \BibitemOpen
  \bibfield  {author} {\bibinfo {author} {\bibfnamefont {L.}~\bibnamefont
  {Dabelow}}, \bibinfo {author} {\bibfnamefont {S.}~\bibnamefont {Bo}}, \ and\
  \bibinfo {author} {\bibfnamefont {R.}~\bibnamefont {Eichhorn}},\ }\href@noop
  {} {\bibfield  {journal} {\bibinfo  {journal} {Physical Review X}\ }\textbf
  {\bibinfo {volume} {9}},\ \bibinfo {pages} {021009} (\bibinfo {year}
  {2019})}\BibitemShut {NoStop}%
\bibitem [{\citenamefont {Speck}(2017)}]{speck2017stochastic}%
  \BibitemOpen
  \bibfield  {author} {\bibinfo {author} {\bibfnamefont {T.}~\bibnamefont
  {Speck}},\ }\href@noop {} {\bibfield  {journal} {\bibinfo  {journal} {arXiv
  preprint arXiv}\ }\textbf {\bibinfo {volume} {1707}} (\bibinfo {year}
  {2017})}\BibitemShut {NoStop}%
\bibitem [{\citenamefont {Speck}(2018)}]{speck2018active}%
  \BibitemOpen
  \bibfield  {author} {\bibinfo {author} {\bibfnamefont {T.}~\bibnamefont
  {Speck}},\ }\href@noop {} {\bibfield  {journal} {\bibinfo  {journal} {EPL
  (Europhysics Letters)}\ }\textbf {\bibinfo {volume} {123}},\ \bibinfo {pages}
  {20007} (\bibinfo {year} {2018})}\BibitemShut {NoStop}%
\bibitem [{\citenamefont {Speck}\ \emph {et~al.}(2015)\citenamefont {Speck},
  \citenamefont {Menzel}, \citenamefont {Bialk{\'e}},\ and\ \citenamefont
  {L{\"o}wen}}]{speck2015dynamical}%
  \BibitemOpen
  \bibfield  {author} {\bibinfo {author} {\bibfnamefont {T.}~\bibnamefont
  {Speck}}, \bibinfo {author} {\bibfnamefont {A.~M.}\ \bibnamefont {Menzel}},
  \bibinfo {author} {\bibfnamefont {J.}~\bibnamefont {Bialk{\'e}}}, \ and\
  \bibinfo {author} {\bibfnamefont {H.}~\bibnamefont {L{\"o}wen}},\ }\href@noop
  {} {\bibfield  {journal} {\bibinfo  {journal} {The Journal of chemical
  physics}\ }\textbf {\bibinfo {volume} {142}},\ \bibinfo {pages} {224109}
  (\bibinfo {year} {2015})}\BibitemShut {NoStop}%
\bibitem [{\citenamefont {GrandPre}\ and\ \citenamefont
  {Limmer}(2018)}]{grandpre2018current}%
  \BibitemOpen
  \bibfield  {author} {\bibinfo {author} {\bibfnamefont {T.}~\bibnamefont
  {GrandPre}}\ and\ \bibinfo {author} {\bibfnamefont {D.~T.}\ \bibnamefont
  {Limmer}},\ }\href@noop {} {\bibfield  {journal} {\bibinfo  {journal}
  {Physical Review E}\ }\textbf {\bibinfo {volume} {98}},\ \bibinfo {pages}
  {060601} (\bibinfo {year} {2018})}\BibitemShut {NoStop}%
\bibitem [{\citenamefont {Giardina}\ \emph {et~al.}(2006)\citenamefont
  {Giardina}, \citenamefont {Kurchan},\ and\ \citenamefont
  {Peliti}}]{giardina2006direct}%
  \BibitemOpen
  \bibfield  {author} {\bibinfo {author} {\bibfnamefont {C.}~\bibnamefont
  {Giardina}}, \bibinfo {author} {\bibfnamefont {J.}~\bibnamefont {Kurchan}}, \
  and\ \bibinfo {author} {\bibfnamefont {L.}~\bibnamefont {Peliti}},\
  }\href@noop {} {\bibfield  {journal} {\bibinfo  {journal} {Physical review
  letters}\ }\textbf {\bibinfo {volume} {96}},\ \bibinfo {pages} {120603}
  (\bibinfo {year} {2006})}\BibitemShut {NoStop}%
\bibitem [{\citenamefont {Ray}\ \emph {et~al.}(2018{\natexlab{a}})\citenamefont
  {Ray}, \citenamefont {Chan},\ and\ \citenamefont {Limmer}}]{ray2018exact}%
  \BibitemOpen
  \bibfield  {author} {\bibinfo {author} {\bibfnamefont {U.}~\bibnamefont
  {Ray}}, \bibinfo {author} {\bibfnamefont {G.~K.-L.}\ \bibnamefont {Chan}}, \
  and\ \bibinfo {author} {\bibfnamefont {D.~T.}\ \bibnamefont {Limmer}},\
  }\href@noop {} {\bibfield  {journal} {\bibinfo  {journal} {Physical review
  letters}\ }\textbf {\bibinfo {volume} {120}},\ \bibinfo {pages} {210602}
  (\bibinfo {year} {2018}{\natexlab{a}})}\BibitemShut {NoStop}%
\bibitem [{\citenamefont {Weeks}\ \emph {et~al.}(1971)\citenamefont {Weeks},
  \citenamefont {Chandler},\ and\ \citenamefont {Andersen}}]{weeks1971role}%
  \BibitemOpen
  \bibfield  {author} {\bibinfo {author} {\bibfnamefont {J.~D.}\ \bibnamefont
  {Weeks}}, \bibinfo {author} {\bibfnamefont {D.}~\bibnamefont {Chandler}}, \
  and\ \bibinfo {author} {\bibfnamefont {H.~C.}\ \bibnamefont {Andersen}},\
  }\href@noop {} {\bibfield  {journal} {\bibinfo  {journal} {The Journal of
  chemical physics}\ }\textbf {\bibinfo {volume} {54}},\ \bibinfo {pages}
  {5237} (\bibinfo {year} {1971})}\BibitemShut {NoStop}%
\bibitem [{\citenamefont {Cagnetta}\ and\ \citenamefont
  {Mallmin}(2020)}]{cagnetta2020efficiency}%
  \BibitemOpen
  \bibfield  {author} {\bibinfo {author} {\bibfnamefont {F.}~\bibnamefont
  {Cagnetta}}\ and\ \bibinfo {author} {\bibfnamefont {E.}~\bibnamefont
  {Mallmin}},\ }\href@noop {} {\bibfield  {journal} {\bibinfo  {journal}
  {Physical Review E}\ }\textbf {\bibinfo {volume} {101}},\ \bibinfo {pages}
  {022130} (\bibinfo {year} {2020})}\BibitemShut {NoStop}%
\bibitem [{\citenamefont {Redner}\ \emph {et~al.}(2013)\citenamefont {Redner},
  \citenamefont {Hagan},\ and\ \citenamefont {Baskaran}}]{redner2013structure}%
  \BibitemOpen
  \bibfield  {author} {\bibinfo {author} {\bibfnamefont {G.~S.}\ \bibnamefont
  {Redner}}, \bibinfo {author} {\bibfnamefont {M.~F.}\ \bibnamefont {Hagan}}, \
  and\ \bibinfo {author} {\bibfnamefont {A.}~\bibnamefont {Baskaran}},\
  }\href@noop {} {\bibfield  {journal} {\bibinfo  {journal} {Physical review
  letters}\ }\textbf {\bibinfo {volume} {110}},\ \bibinfo {pages} {055701}
  (\bibinfo {year} {2013})}\BibitemShut {NoStop}%
\bibitem [{\citenamefont {Cates}\ and\ \citenamefont
  {Tailleur}(2015)}]{cates2015motility}%
  \BibitemOpen
  \bibfield  {author} {\bibinfo {author} {\bibfnamefont {M.~E.}\ \bibnamefont
  {Cates}}\ and\ \bibinfo {author} {\bibfnamefont {J.}~\bibnamefont
  {Tailleur}},\ }\href@noop {} {\bibfield  {journal} {\bibinfo  {journal}
  {Annu. Rev. Condens. Matter Phys.}\ }\textbf {\bibinfo {volume} {6}},\
  \bibinfo {pages} {219} (\bibinfo {year} {2015})}\BibitemShut {NoStop}%
\bibitem [{\citenamefont {Chiarantoni}\ \emph {et~al.}(2020)\citenamefont
  {Chiarantoni}, \citenamefont {Cagnetta}, \citenamefont {Corberi},
  \citenamefont {Gonnella},\ and\ \citenamefont {Suma}}]{chiarantoni2020work}%
  \BibitemOpen
  \bibfield  {author} {\bibinfo {author} {\bibfnamefont {P.}~\bibnamefont
  {Chiarantoni}}, \bibinfo {author} {\bibfnamefont {F.}~\bibnamefont
  {Cagnetta}}, \bibinfo {author} {\bibfnamefont {F.}~\bibnamefont {Corberi}},
  \bibinfo {author} {\bibfnamefont {G.}~\bibnamefont {Gonnella}}, \ and\
  \bibinfo {author} {\bibfnamefont {A.}~\bibnamefont {Suma}},\ }\href@noop {}
  {\bibfield  {journal} {\bibinfo  {journal} {Journal of Physics A:
  Mathematical and Theoretical}\ } (\bibinfo {year} {2020})}\BibitemShut
  {NoStop}%
\bibitem [{\citenamefont {Ray}\ \emph {et~al.}(2018{\natexlab{b}})\citenamefont
  {Ray}, \citenamefont {Chan},\ and\ \citenamefont
  {Limmer}}]{ray2018importance}%
  \BibitemOpen
  \bibfield  {author} {\bibinfo {author} {\bibfnamefont {U.}~\bibnamefont
  {Ray}}, \bibinfo {author} {\bibfnamefont {G.~K.-L.}\ \bibnamefont {Chan}}, \
  and\ \bibinfo {author} {\bibfnamefont {D.~T.}\ \bibnamefont {Limmer}},\
  }\href@noop {} {\bibfield  {journal} {\bibinfo  {journal} {The Journal of
  chemical physics}\ }\textbf {\bibinfo {volume} {148}},\ \bibinfo {pages}
  {124120} (\bibinfo {year} {2018}{\natexlab{b}})}\BibitemShut {NoStop}%
\bibitem [{\citenamefont {Dean}(1996)}]{dean1996langevin}%
  \BibitemOpen
  \bibfield  {author} {\bibinfo {author} {\bibfnamefont {D.~S.}\ \bibnamefont
  {Dean}},\ }\href@noop {} {\bibfield  {journal} {\bibinfo  {journal} {Journal
  of Physics A: Mathematical and General}\ }\textbf {\bibinfo {volume} {29}},\
  \bibinfo {pages} {L613} (\bibinfo {year} {1996})}\BibitemShut {NoStop}%
\bibitem [{\citenamefont {Tailleur}\ and\ \citenamefont
  {Cates}(2008)}]{tailleur2008statistical}%
  \BibitemOpen
  \bibfield  {author} {\bibinfo {author} {\bibfnamefont {J.}~\bibnamefont
  {Tailleur}}\ and\ \bibinfo {author} {\bibfnamefont {M.}~\bibnamefont
  {Cates}},\ }\href@noop {} {\bibfield  {journal} {\bibinfo  {journal}
  {Physical review letters}\ }\textbf {\bibinfo {volume} {100}},\ \bibinfo
  {pages} {218103} (\bibinfo {year} {2008})}\BibitemShut {NoStop}%
\bibitem [{\citenamefont {Chakraborti}\ \emph {et~al.}(2016)\citenamefont
  {Chakraborti}, \citenamefont {Mishra},\ and\ \citenamefont
  {Pradhan}}]{chakraborti2016additivity}%
  \BibitemOpen
  \bibfield  {author} {\bibinfo {author} {\bibfnamefont {S.}~\bibnamefont
  {Chakraborti}}, \bibinfo {author} {\bibfnamefont {S.}~\bibnamefont {Mishra}},
  \ and\ \bibinfo {author} {\bibfnamefont {P.}~\bibnamefont {Pradhan}},\
  }\href@noop {} {\bibfield  {journal} {\bibinfo  {journal} {Physical Review
  E}\ }\textbf {\bibinfo {volume} {93}},\ \bibinfo {pages} {052606} (\bibinfo
  {year} {2016})}\BibitemShut {NoStop}%
\bibitem [{\citenamefont {Dolezal}\ and\ \citenamefont
  {Jack}(2019)}]{dolezal2019large}%
  \BibitemOpen
  \bibfield  {author} {\bibinfo {author} {\bibfnamefont {J.}~\bibnamefont
  {Dolezal}}\ and\ \bibinfo {author} {\bibfnamefont {R.~L.}\ \bibnamefont
  {Jack}},\ }\href@noop {} {\bibfield  {journal} {\bibinfo  {journal} {Journal
  of Statistical Mechanics: Theory and Experiment}\ }\textbf {\bibinfo {volume}
  {2019}},\ \bibinfo {pages} {123208} (\bibinfo {year} {2019})}\BibitemShut
  {NoStop}%
\bibitem [{\citenamefont {Das}\ and\ \citenamefont
  {Limmer}(2019)}]{das2019variational}%
  \BibitemOpen
  \bibfield  {author} {\bibinfo {author} {\bibfnamefont {A.}~\bibnamefont
  {Das}}\ and\ \bibinfo {author} {\bibfnamefont {D.~T.}\ \bibnamefont
  {Limmer}},\ }\href@noop {} {\bibfield  {journal} {\bibinfo  {journal} {The
  Journal of Chemical Physics}\ }\textbf {\bibinfo {volume} {151}},\ \bibinfo
  {pages} {244123} (\bibinfo {year} {2019})}\BibitemShut {NoStop}%
\bibitem [{\citenamefont {Tociu}\ \emph {et~al.}(2019)\citenamefont {Tociu},
  \citenamefont {Fodor}, \citenamefont {Nemoto},\ and\ \citenamefont
  {Vaikuntanathan}}]{tociu2019dissipation}%
  \BibitemOpen
  \bibfield  {author} {\bibinfo {author} {\bibfnamefont {L.}~\bibnamefont
  {Tociu}}, \bibinfo {author} {\bibfnamefont {{\'E}.}~\bibnamefont {Fodor}},
  \bibinfo {author} {\bibfnamefont {T.}~\bibnamefont {Nemoto}}, \ and\ \bibinfo
  {author} {\bibfnamefont {S.}~\bibnamefont {Vaikuntanathan}},\ }\href@noop {}
  {\bibfield  {journal} {\bibinfo  {journal} {Physical Review X}\ }\textbf
  {\bibinfo {volume} {9}},\ \bibinfo {pages} {041026} (\bibinfo {year}
  {2019})}\BibitemShut {NoStop}%
\bibitem [{\citenamefont {Crosato}\ \emph {et~al.}(2019)\citenamefont
  {Crosato}, \citenamefont {Prokopenko},\ and\ \citenamefont
  {Spinney}}]{crosato2019irreversibility}%
  \BibitemOpen
  \bibfield  {author} {\bibinfo {author} {\bibfnamefont {E.}~\bibnamefont
  {Crosato}}, \bibinfo {author} {\bibfnamefont {M.}~\bibnamefont {Prokopenko}},
  \ and\ \bibinfo {author} {\bibfnamefont {R.~E.}\ \bibnamefont {Spinney}},\
  }\href@noop {} {\bibfield  {journal} {\bibinfo  {journal} {Physical Review
  E}\ }\textbf {\bibinfo {volume} {100}},\ \bibinfo {pages} {042613} (\bibinfo
  {year} {2019})}\BibitemShut {NoStop}%
\bibitem [{\citenamefont {Nigmatullin}\ and\ \citenamefont
  {Prokopenko}(2019)}]{nigmatullin2019thermodynamic}%
  \BibitemOpen
  \bibfield  {author} {\bibinfo {author} {\bibfnamefont {R.}~\bibnamefont
  {Nigmatullin}}\ and\ \bibinfo {author} {\bibfnamefont {M.}~\bibnamefont
  {Prokopenko}},\ }\href@noop {} {\bibfield  {journal} {\bibinfo  {journal}
  {arXiv:1912.08948}\ } (\bibinfo {year} {2019})}\BibitemShut {NoStop}%
\bibitem [{\citenamefont {Solon}\ \emph
  {et~al.}(2015{\natexlab{a}})\citenamefont {Solon}, \citenamefont
  {Stenhammar}, \citenamefont {Wittkowski}, \citenamefont {Kardar},
  \citenamefont {Kafri}, \citenamefont {Cates},\ and\ \citenamefont
  {Tailleur}}]{solon2015pressure}%
  \BibitemOpen
  \bibfield  {author} {\bibinfo {author} {\bibfnamefont {A.~P.}\ \bibnamefont
  {Solon}}, \bibinfo {author} {\bibfnamefont {J.}~\bibnamefont {Stenhammar}},
  \bibinfo {author} {\bibfnamefont {R.}~\bibnamefont {Wittkowski}}, \bibinfo
  {author} {\bibfnamefont {M.}~\bibnamefont {Kardar}}, \bibinfo {author}
  {\bibfnamefont {Y.}~\bibnamefont {Kafri}}, \bibinfo {author} {\bibfnamefont
  {M.~E.}\ \bibnamefont {Cates}}, \ and\ \bibinfo {author} {\bibfnamefont
  {J.}~\bibnamefont {Tailleur}},\ }\href@noop {} {\bibfield  {journal}
  {\bibinfo  {journal} {Physical review letters}\ }\textbf {\bibinfo {volume}
  {114}},\ \bibinfo {pages} {198301} (\bibinfo {year}
  {2015}{\natexlab{a}})}\BibitemShut {NoStop}%
\bibitem [{\citenamefont {Fily}\ \emph {et~al.}(2017)\citenamefont {Fily},
  \citenamefont {Baskaran},\ and\ \citenamefont {Hagan}}]{fily2017equilibrium}%
  \BibitemOpen
  \bibfield  {author} {\bibinfo {author} {\bibfnamefont {Y.}~\bibnamefont
  {Fily}}, \bibinfo {author} {\bibfnamefont {A.}~\bibnamefont {Baskaran}}, \
  and\ \bibinfo {author} {\bibfnamefont {M.~F.}\ \bibnamefont {Hagan}},\
  }\href@noop {} {\bibfield  {journal} {\bibinfo  {journal} {The European
  Physical Journal E}\ }\textbf {\bibinfo {volume} {40}},\ \bibinfo {pages} {1}
  (\bibinfo {year} {2017})}\BibitemShut {NoStop}%
\bibitem [{\citenamefont {L{\"o}wen}(2018)}]{lowen2018active}%
  \BibitemOpen
  \bibfield  {author} {\bibinfo {author} {\bibfnamefont {H.}~\bibnamefont
  {L{\"o}wen}},\ }\href@noop {} {\bibfield  {journal} {\bibinfo  {journal} {EPL
  (Europhysics Letters)}\ }\textbf {\bibinfo {volume} {121}},\ \bibinfo {pages}
  {58001} (\bibinfo {year} {2018})}\BibitemShut {NoStop}%
\bibitem [{\citenamefont {Winkler}(2016)}]{winkler2016dynamics}%
  \BibitemOpen
  \bibfield  {author} {\bibinfo {author} {\bibfnamefont {R.~G.}\ \bibnamefont
  {Winkler}},\ }\href@noop {} {\bibfield  {journal} {\bibinfo  {journal} {Soft
  matter}\ }\textbf {\bibinfo {volume} {12}},\ \bibinfo {pages} {3737}
  (\bibinfo {year} {2016})}\BibitemShut {NoStop}%
\bibitem [{\citenamefont {Hidalgo}\ \emph {et~al.}(2017)\citenamefont
  {Hidalgo}, \citenamefont {Nemoto},\ and\ \citenamefont
  {Lecomte}}]{hidalgo2017finite}%
  \BibitemOpen
  \bibfield  {author} {\bibinfo {author} {\bibfnamefont {E.~G.}\ \bibnamefont
  {Hidalgo}}, \bibinfo {author} {\bibfnamefont {T.}~\bibnamefont {Nemoto}}, \
  and\ \bibinfo {author} {\bibfnamefont {V.}~\bibnamefont {Lecomte}},\
  }\href@noop {} {\bibfield  {journal} {\bibinfo  {journal} {Physical Review
  E}\ }\textbf {\bibinfo {volume} {95}},\ \bibinfo {pages} {062134} (\bibinfo
  {year} {2017})}\BibitemShut {NoStop}%
\bibitem [{\citenamefont {Takatori}\ \emph {et~al.}(2014)\citenamefont
  {Takatori}, \citenamefont {Yan},\ and\ \citenamefont
  {Brady}}]{takatori2014swim}%
  \BibitemOpen
  \bibfield  {author} {\bibinfo {author} {\bibfnamefont {S.~C.}\ \bibnamefont
  {Takatori}}, \bibinfo {author} {\bibfnamefont {W.}~\bibnamefont {Yan}}, \
  and\ \bibinfo {author} {\bibfnamefont {J.~F.}\ \bibnamefont {Brady}},\
  }\href@noop {} {\bibfield  {journal} {\bibinfo  {journal} {Physical review
  letters}\ }\textbf {\bibinfo {volume} {113}},\ \bibinfo {pages} {028103}
  (\bibinfo {year} {2014})}\BibitemShut {NoStop}%
\bibitem [{\citenamefont {Omar}\ \emph {et~al.}(2020)\citenamefont {Omar},
  \citenamefont {Wang},\ and\ \citenamefont {Brady}}]{omar2020microscopic}%
  \BibitemOpen
  \bibfield  {author} {\bibinfo {author} {\bibfnamefont {A.~K.}\ \bibnamefont
  {Omar}}, \bibinfo {author} {\bibfnamefont {Z.-G.}\ \bibnamefont {Wang}}, \
  and\ \bibinfo {author} {\bibfnamefont {J.~F.}\ \bibnamefont {Brady}},\
  }\href@noop {} {\bibfield  {journal} {\bibinfo  {journal} {Physical Review
  E}\ }\textbf {\bibinfo {volume} {101}},\ \bibinfo {pages} {012604} (\bibinfo
  {year} {2020})}\BibitemShut {NoStop}%
\bibitem [{\citenamefont {Touchette}(2018)}]{touchette2018introduction}%
  \BibitemOpen
  \bibfield  {author} {\bibinfo {author} {\bibfnamefont {H.}~\bibnamefont
  {Touchette}},\ }\href@noop {} {\bibfield  {journal} {\bibinfo  {journal}
  {Physica A: Statistical Mechanics and its Applications}\ }\textbf {\bibinfo
  {volume} {504}},\ \bibinfo {pages} {5} (\bibinfo {year} {2018})}\BibitemShut
  {NoStop}%
\bibitem [{\citenamefont {Chetrite}\ and\ \citenamefont
  {Touchette}(2015{\natexlab{b}})}]{chetrite2015variational}%
  \BibitemOpen
  \bibfield  {author} {\bibinfo {author} {\bibfnamefont {R.}~\bibnamefont
  {Chetrite}}\ and\ \bibinfo {author} {\bibfnamefont {H.}~\bibnamefont
  {Touchette}},\ }\href@noop {} {\bibfield  {journal} {\bibinfo  {journal}
  {Journal of Statistical Mechanics: Theory and Experiment}\ }\textbf {\bibinfo
  {volume} {2015}},\ \bibinfo {pages} {P12001} (\bibinfo {year}
  {2015}{\natexlab{b}})}\BibitemShut {NoStop}%
\bibitem [{\citenamefont {Solon}\ \emph
  {et~al.}(2015{\natexlab{b}})\citenamefont {Solon}, \citenamefont {Cates},\
  and\ \citenamefont {Tailleur}}]{solon2015active}%
  \BibitemOpen
  \bibfield  {author} {\bibinfo {author} {\bibfnamefont {A.~P.}\ \bibnamefont
  {Solon}}, \bibinfo {author} {\bibfnamefont {M.}~\bibnamefont {Cates}}, \ and\
  \bibinfo {author} {\bibfnamefont {J.}~\bibnamefont {Tailleur}},\ }\href@noop
  {} {\bibfield  {journal} {\bibinfo  {journal} {The European Physical Journal
  Special Topics}\ }\textbf {\bibinfo {volume} {224}},\ \bibinfo {pages} {1231}
  (\bibinfo {year} {2015}{\natexlab{b}})}\BibitemShut {NoStop}%
\bibitem [{\citenamefont {Bialk{\'e}}\ \emph {et~al.}(2013)\citenamefont
  {Bialk{\'e}}, \citenamefont {L{\"o}wen},\ and\ \citenamefont
  {Speck}}]{bialke2013microscopic}%
  \BibitemOpen
  \bibfield  {author} {\bibinfo {author} {\bibfnamefont {J.}~\bibnamefont
  {Bialk{\'e}}}, \bibinfo {author} {\bibfnamefont {H.}~\bibnamefont
  {L{\"o}wen}}, \ and\ \bibinfo {author} {\bibfnamefont {T.}~\bibnamefont
  {Speck}},\ }\href@noop {} {\bibfield  {journal} {\bibinfo  {journal} {EPL
  (Europhysics Letters)}\ }\textbf {\bibinfo {volume} {103}},\ \bibinfo {pages}
  {30008} (\bibinfo {year} {2013})}\BibitemShut {NoStop}%
\bibitem [{\citenamefont {Angeletti}\ and\ \citenamefont
  {Touchette}(2016)}]{angeletti2016diffusions}%
  \BibitemOpen
  \bibfield  {author} {\bibinfo {author} {\bibfnamefont {F.}~\bibnamefont
  {Angeletti}}\ and\ \bibinfo {author} {\bibfnamefont {H.}~\bibnamefont
  {Touchette}},\ }\href@noop {} {\bibfield  {journal} {\bibinfo  {journal}
  {Journal of Mathematical Physics}\ }\textbf {\bibinfo {volume} {57}},\
  \bibinfo {pages} {023303} (\bibinfo {year} {2016})}\BibitemShut {NoStop}%
\end{thebibliography}
\end{document}